\documentclass[preprint,10pt]{sigplanconf}

\usepackage{amsmath}
\usepackage[pdftex]{graphicx}
\usepackage{multirow}
\usepackage[table]{xcolor}

\begin{document}

\setlength{\pdfpageheight}{\paperheight}
\setlength{\pdfpagewidth}{\paperwidth}

\conferenceinfo{OOPSLA '15}{October 25--30, 2015, Pittsburgh, PA, USA} 
\copyrightyear{2015} 
\copyrightdata{978-1-nnnn-nnnn-n/yy/mm} 
\doi{nnnnnnn.nnnnnnn}

% Uncomment one of the following two, if you are not going for the 
% traditional copyright transfer agreement.

%\exclusivelicense                % ACM gets exclusive license to publish, 
                                  % you retain copyright

%\permissiontopublish             % ACM gets nonexclusive license to publish
                                  % (paid open-access papers, 
                                  % short abstracts)

\titlebanner{Preprint}        % These are ignored unless
\preprintfooter{Second phase submission to OOPSLA'15}   % 'preprint' option specified.

\title{How Scale Affects Structure in Java Programs}
%\subtitle{Subtitle Text, if any}

%% \authorinfo{}
%%            {}
%%            {}

\authorinfo{Cristina V. Lopes}
           {Bren School of Information and Computer Sciences\\
            University of California, Irvine, USA}
           {lopes@uci.edu}
\authorinfo{Joel Ossher}
           {Bren School of Information and Computer Sciences\\
            University of California, Irvine, USA}
           {ossher@gmail.com}

\maketitle

\begin{abstract}
Many internal software metrics and external quality attributes of Java
programs correlate strongly with program size. This knowledge has been
used pervasively in quantitative studies of software through practices
such as normalization on size metrics. This paper reports size-related
super- and sublinear effects that have not been known before. Findings
obtained on a very large collection of Java programs -- 30,911
projects hosted at Google Code as of Summer 2011 -- unveils how
certain characteristics of programs vary {\em disproportionately} with
program size, sometimes even non-monotonically. Many of the specific
parameters of nonlinear relations are reported. This result gives
further insights for the differences of ``programming in the small''
vs. ``programming in the large.''  The reported findings carry
important consequences for OO software metrics, and software research
in general: metrics that have been known to correlate with size can
now be properly normalized so that all the information that is left in
them is size-independent.

\end{abstract}

\category{Software and its engineering}{Software organization and properties}{Software system structures}

% general terms are not compulsory anymore, 
% you may leave them out
%\terms
%term1, term2


\section{Introduction}

\begin{figure}
\centering
\includegraphics[width=1.3in]{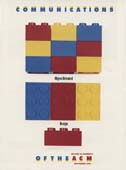}
\includegraphics[width=1.8in]{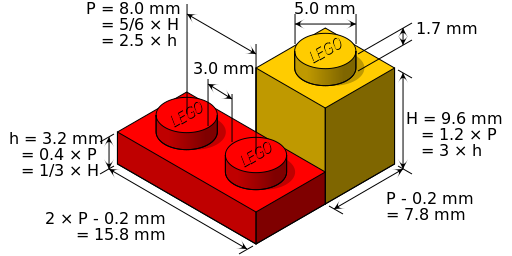}
\caption{Left: Cover of the CACM, Special issue on Object-Oriented
  Programming, September 1990~\cite{CACM:1990}. Right: LEGO bricks
  showing standard dimensions. (Source: Wikipedia, ``Cmglee''.)}
\label{fig:CACM}
\end{figure}

Early on in the history of programming, a metaphor was put forward
that has seen wide acceptance in the software community: that of
programming as LEGO (Figure~\ref{fig:CACM}). The metaphor suggests
that building large systems is a matter of connecting small
standardized bricks together, one at a time, through their universal
interfaces: the small bricks are independent of the scale and purpose
of the construction. This metaphor had a tremendous influence
in the development of OOP languages. Inspired by the simplicity of the
LEGO construction model, these languages placed their focus on
mechanisms that would allow to connect small computational units
together to create large software systems.

Meanwhile, in 1975 another idea was put forward that has also seen
wide acceptance in the software community: that ``programming in the
large'' has different characteristics from ``programming in the
small.'' This idea was first formulated by DeRemer and
Kron~\cite{DeRemer:1975}, who argued that ``structuring a large
collection of modules to form a "system" is an essentially distinct
and different intellectual activity from that of constructing the
individual modules.'' DeRemer and Kron went on to advocate a ``Module
Interconnection Language'' (MIL) for large systems.

These two popular ideas aren't mutually exclusive: it is possible to
imagine system-wide directives and constraints (i.e. architecture) for
large LEGO constructions. But DeRemer and Kron's essay states a
premise that puts some pressure on the LEGO metaphor: ``Where an MIL
is not available, module interconnectivity information is usually
buried partly in the modules, partly in an often amorphous collection
of linkage-editor instructions, and partly in the informal
documentation of the project.'' In LEGO terms, this might mean that in
order to build a large castle, one might need to plumb stronger
connection material into the bricks themselves. In short, the scale of
the system would affect the internal structure of the construction.

This paper focuses on the core of these two popular ideas by asking
and answering the following question:

\noindent
{\bf Does the scale of the software system affect the internal
  structure of its modules or are modules scale-invariant?}

We want to find out whether there are mathematical principles
related to size in large ecosystems of software projects. Besides
shedding light on the differences between programming-in-the-small and
programming-in-the-large, this question has important
  implications for research. A common practice for validating ideas
in software research is to collect a number of artifacts, either
randomly or using some criteria, measure the effects of the ideas
using those artifacts, and reach conclusions from the empirical
data. Even though size of software artifacts (projects, classes, etc.)
has been known to be an issue in quantitative studies of software,
software research continues to be fairly oblivious to its effect in
these assorted datasets. This is particularly problematic for any
studies involving software metrics, including OO metrics. It also
affects performance studies that tend to collect data on relatively
small programs that aren't necessarily representative of large
programs.  Several studies published in the literature may have
reached invalid conclusions by ignoring the effect of size or by
treating it inappropriately.

The question, as formulated above, is too ambitious to be answered in
one single step. This paper takes only the first step. We focus on
Object-Oriented software systems, since those are the most influenced
by the programming-as-LEGO metaphor; other language families should be
studied for broader conclusions. Within OOP, we focus on Java, since
it is one of the most popular OOP languages; other OOP ecosystems
should be studied for broader conclusions. Finally, we report on a
dozen metrics that illustrate the main trends, but many more metrics
could be studied.

We deconstruct the general question into five research questions for
which specific metrics can be measured:

\begin{itemize}
\item [RQ1] Module size: Are modules of larger systems larger than modules
  of smaller systems?
\item [RQ2] Module Type: Is there a statistically significant variation
  in the mix of classes and interfaces for projects of different size
  scales?
\item [RQ3] Internal Complexity: Are modules of larger systems more,
  or fewer, complex than modules of smaller systems?
\item [RQ4] Composition via Inheritance: Does the scale of the project
affect the use of inheritance?
\item [RQ5] Dependencies: Do larger projects use disproportionately 
  more, or fewer, types from external libraries than smaller projects?
\end{itemize}

This study puts forward strong evidence that, as programs become
larger, the internal structure of the modules and the mixture of
composition mechanisms used are affected. As such, the paper
makes the following contributions:

\begin{enumerate}
\item It unveils strong empirical evidence of the existence of super-
  and sublinear effects in software that have not been measured
  before, and it shows concrete parameters of many non-linear
  relations that underly a large and important ecosystem of Java
  programs.
\item It proposes more accurate definitions of popular OO metrics that
  properly normalize for size.
\item By unveilling the characteristics of large projects, it may
  suggest new ideas for how to tame deterimental non-linear effects,
  both in terms of programming language design and project management.
\end{enumerate}

\section{Motivation and Related Work}
\label{sec:relwork}

It has been almost 25 years since Chidamber and Kemerer published
their influential paper on OO metrics at
OOPSLA'91~\cite{Chidamber:1991}. Since then, OO metrics have been used
pervasively in research and development. Here, we review and
discuss the main issues with OO metrics, and the research
community's attempts to understand the empirically-based principles of
software.

\subsection{The Confusing Effect of Size}
\label{sec:confusingeffect}

A large body of literature exists in analyzing how software metrics
correlate with software quality. A typical study along those lines
involves computing internal software metrics (e.g. coupling of
classes) and correlating them with external quality attributes
(e.g. post-release bug fixes involving those classes). Many studies of
this kind apply simple univariate statistical analysis, and often
conclude that there is a correlation.

For quite some time, however, size has been known to be a potential
confounding factor in empirical studies of software artifacts. For
example, in a study designed to verify whether it is possible to use a
multivariate logistic regression model based on OO metrics to predict
faults in OO programs, Briand et al.~\cite{Briand2000245} reported
strong correlations between class size and several OO software
metrics. They then went on to compensate for that correlation by doing
partial correlations. In another study of a large C++
system~\cite{Cartwright2000}, Cartwright et al. also reported such
correlations. In 2001, El Emam et al.~\cite{ElEmam2001} presented a
comprehensive analysis of the effect of class size in several OO
metrics, and suggested that this effect might have confounded prior
studies.\footnote{We refer readers to ~\cite{ElEmam2001} for an
  extensive list of studies that the authors suggest may have reached
  invalid conclusions by neglecting to compensate for size.}  They
then presented their own study of a large C++ framework which showed
that strong correlations resulting from univariate analysis of data
were neutralized when multivariate analysis including class size is
used. Another more recent study reached the same conclusions when
studying the relation between internal software attributes and
component utilization~\cite{Sajnani2014}.

However, Briand et al. and El Emam et al.'s argument has drawn some
criticism stemming from the point of view that multivariate analysis
of the kind proposed in their papers produces ill-specified, logically
inconsistent statistical models~\cite{Evanco2003}. Specifically, the
partial correlation of $X$ and $Y$ controlling for a third variable
$Z$, written $r(X,Y|Z)$, is a measure of the relationship between X
and Y if statistically {\em we hold Z constant}. But trying to
predict, for example, the effect on post-release defects $X$ by
increasing the coupling value $Y$ while holding the number of lines of
code $Z$ constant doesn't make sense, because in the world from where
the data comes, increasing coupling usually requires additional lines
of code (e.g. field and variable declarations). As Evanco points
out~\cite{Evanco2003}, this model is inconsistent with the reality of
the data. The suggestion following the criticism is that prediction
models should use the metric in question $Y$ or the size metric ($Z$),
whichever gives more predictive power, but not both.

Either way, these observations raise doubts about the value of the
many software metrics that are correlated with size, as they do not
provide any more additional statistical power than what is already
provided by their strong correlate -- and size is very easy to
measure. 
In summary, size may not be a confounding factor in statistical
terminology, but it certainly has been the source of much confusion in
software research.

\subsection{Non-Normal Data}
\label{sec:nonnormal}

In their study of slice-based cohesion and coupling metrics over 63 C
programs, Meyers and Binkley ~\cite{Meyers:2007} include correlation
coefficients between several coupling and cohesion metrics and Lines
of Code (LOC). They show that they are not correlated. We noted that
their correlation analysis was made for the entire dataset, which
contained components of considerably different sizes; this made the
analysis prone to sknewness-related errors. In subsequent email
exchanges with one of the authors, he kindly shared the data with us;
we then verified that, indeed, the distribution of size of the
components was not normal but log-normal. Once the transformation to
log scale was performed, the data showed moderate-to-strong positive
linear correlation between log(size) and their coupling
metric.

This exchange illustrates another source of problems when doing empirical
studies of software artifacts, and how size can drastically affect the
conclusions. Size is not just a confusing factor; because the
projects' size distribution is often skewed, the statistical analysis
needs to take non-normal data into account too.

\subsection{Software Corpora}
\label{sec:corpora}

In recent years, there has been an increasing number of empirical
studies on increasingly larger collections of software projects for
purposes of understanding the way that developers use programming
languages in real projects. 
For example, Tempero et
al.~\cite{Tempero:2008} studied the way Java programs use inheritance
in the 100 projects of the Qualitas
corpus~\cite{QualitasCorpus:APSEC:2010}. The criteria for inclusion of
projects in that corpus is relatively strict, requiring, for example,
distribution in both source and binary forms.\footnote{See
  https://www.cs.auckland.ac.nz/$\sim$ewan/corpus/docs/criteria.html} While
their findings fall within the results reported here, the Qualitas
corpus contains only 100 projects. The results reported in
\cite{Tempero:2008} show that the data does not follow a normal
distribution. Another study on the same corpus explored the simulated
use of multiple dispatch via cascading {\texttt instanceof}
statements~\cite{Muschevici:2008}. Another study by Gil and Lenz
\cite{Gil:2010} studied the use of overloading in Java programs, also
using the Qualitas corpus. Some of the conclusions in these studies
(e.g. whether a project is an outlier or not) may be missing the
effect of size of the project.

Calla\'u et al.~\cite{Callau:2011} made a statistical analysis of
1,000 Smalltalk projects found in SqueakSource in order to understand
the use of certain dynamic features of Smalltalk. They do not
report the distribution in terms of project size. The study was
designed to gather bulk statistics along an existing taxonomy, so the
results are reported as simple counts of feature occurrences among the
whole corpus or among a category of projects (e.g. out of 652,990
methods, only 8,349 use dynamic features, and then a breakdown is
shown among categories). While the taxonomy is taken into account in the
analysis of the data, project size is not. It would be interesting
to see whether there is a correlation between the categories and size
of the projects.

Collberg et al.~\cite{Collberg:2007} randomly collected 1,132 jar
files off the Internet and analyzed them (at bytecode level) using a
tool developed by the authors. The purpose of that study was to inform
Java language designers and implementers about how developers actually
use the language. That study reports summary statistics for their
entire dataset without taking the distribution of jar size into
account. Most distributions shown in the paper aren't normal, so the
summary statistics are somewhat misleading. Some of the reported
metrics in that study are the same metrics that we use for our study;
for example they found on average 9 methods per class, with median 5.
The reported values fall within the range of ours, but particularly
close to the values for large projects, which leads us to believe that
their dataset was biased towards large projects.

In another large study, Grechanik et al.~\cite{Grechanik:2010} have
conducted an empirical assessment of 2,080 Java projects randomly
selected from Sourceforge, and discovered several facts about the
projects' use of Java. The size of the projects is not reported, and
only simple statistics are given. For example, the reported mean and
median methods per class are 3.5 and 4, respectively. Given that the
data does not follow a normal distribution on project size, these
values are, again, somewhat misleading and at odds with the findings
of Collberg et al.~\cite{Collberg:2007}. Like so many large open source
code repositories, Sourceforge is severely skewed towards small to
medium projects; the reported summary statistics are consistent with
our findings for small projects.

In any large corpora of projects, the data rarely follows normal
distributions of size, so simple summary statistics such as averages
and medians reported in some of these papers provide only weak
insights into the principles of those ecosystems, and may hide
important phenomena. Also, sample biases may have a large influence on
assumptions and conclusions.
But what exactly is the effect of size on software artifacts? Can we
find general statistical principles that explain the 
phenomena observed in prior studies? 

\subsection{Complex Systems}

Ours is not the first study to try to unveil internal mathematical
structures of software, and the software research community is not the
only one looking for mathematical principles in existing software;
communities that study complex systems and networks have long found
software intriguing. One of the first studies of this kind was by
Valverde et al.~\cite{Valverde:2002}, which analyzed the types and
dependencies in the JDK, and noticed the existence of power laws and
small world behavior. Soon after, Myers~\cite{Myers:2003} explored
what he called ``collaboration graphs'' ({\em aka} dependencies) in
three C++ and three C applications. Many more studies of this kind
followed. For example, \cite{Valverde:2005}, \cite{Zheng:2008},
\cite{Fortuna:2011} and \cite{Gherardi:2013} all study the evolution
of software networks finding evidence of known mathematical principles
that also exist in natural systems, and that might serve as predictive
models for software evolution.

Closer to our work, a study presented in 2006 by Baxter et
al.~\cite{Baxter:2006} also targeted the ``Lego Hypothesis,'' as
coined by the authors. That study, which built on an earlier one by
the same group~\cite{Potanin:2005}, searched for the existence of
power laws and other mathematical functions in a collection of 56 Java
applications using 17 OO metrics, such as number of methods per type
and the number of dependencies per type. For each of those 56
applications, the study revealed whether the 17 metrics' data points
could fit the mathematical functions of interest. The study found that
very few projects, and in only very few metrics, had strict power law
distributions; most projects, and in most metrics, revealed reasonable
fits at 80\% confidence interval with several of the functions that
they were searching for. Another study by Louridas et
al.~\cite{Louridas:2008} studied the existence of power laws in a
variety of applications written in a variety of languages.

All of these studies largely ignore {\em application} size, and focus
on the modules themselves (i.e. classes, interfaces). In the study by
Baxter et al.~\cite{Baxter:2006}, the results are ordered by
application size, and even grouped within size ranges; but no insights
are given regarding the effect, if any, that application size may have
on the observations. We believe our study is complementary to all of
these prior studies in search for mathematical laws in software
applications, because it focuses on the size of the application as a
whole, not just on the size of each OO module.

\section{Dataset}
\label{sec:dataset}

In this study, we use the Sourcerer 2011
dataset~~\cite{Lopes+Ossher:2012}, which contains over 150,000+
projects collected from Google Code, SourceForge and Apache as of
2011. The projects have been processed into a relational database of
entities and relations, using the Sourcerer Tools publicly available
from Github~\cite{Sourcerer:2015}. The database facilitates static
analysis for very large collections of source code, as it contains
preprocessed static analysis information that can be queried on
demand. By issuing specific queries on the database, we extracted the
necessary numbers into a Comma Separated Value (CSV) file, which was
then used to perform the statistical analysis described in this paper.

The database produced by the Sourcerer tools was, therefore, the basis
of our study. We present a small example that illustrates the kinds of
entities and relations that are found in the database. Consider the
following Java program:

{\footnotesize
\begin{verbatim}
package foo;

public class FooNumber {
  private int x;
  FooNumber(int _x) { x = _x; }
  private void print() {
    System.out.println("It is number " + x)
  }
  public static void main(String[] args) {
    new FooNumber(Integer.parseInt(args[0])).print();
  }
}
\end{verbatim}
}

\begin{table}
\centering
\caption{Entities}
\label{tab:entities}
{\small
\begin{tabular}{|l|l|l|} \hline
Entity ID & FQN & Type\\ \hline
1 & foo & PACKAGE \\
2 & foo.FooNumber & CLASS \\
3 & foo.FooNumber.x & FIELD \\
4 & foo.FooNumber.$<$init$>$ & CONSTRUCTOR \\
5 & foo.FooNumber.print & METHOD \\
6 & foo.FooNumber.main & METHOD \\ 
... & ... & ... \\\hline
\end{tabular}\\

\caption{Relations}
\label{tab:relations}
\begin{tabular}{|l|c|l|} \hline
Source & Relation type & Target \\ \hline
1 & CONTAINS & 2 \\
2 & CONTAINS & 3 \\
2 & CONTAINS & 4 \\
2 & CONTAINS & 5 \\
2 & CONTAINS & 6 \\
3 & HOLDS  & {\em Integer\_ID} \\
4 & WRITES & 3 \\
5 & READS  & 3 \\
5 & CALLS  & {\em println\_ID} \\
6 & INSTANTIATES & 4 \\
6 & CALLS & 5 \\ 
... & ... & ... \\\hline
\end{tabular}
}
\end{table}

This program results in the entities and relations shown in
Tables~\ref{tab:entities} and \ref{tab:relations} (not all
entities and relations are shown, for brevity sake).
Given this database schema, with these entities and relations tables,
we issued several queries in order to extract all the numbers we
needed. Here is one example query that extracts the number of methods
declared in classes in each of the projects:

{\footnotesize
\begin{verbatim}
-- Extract number of class methods per project
SELECT p.project_id,IFNULL(COUNT(DISTINCT m.entity_id),0) 
  FROM e_methods AS m
  INNER JOIN r_contains AS r ON m.entity_id = r.rhs_eid
  INNER JOIN e_classes AS c ON c.entity_id = r.lhs_eid
  RIGHT JOIN projects AS p ON p.project_id=m.project_id
  GROUP BY p.project_id
\end{verbatim}
}

\begin{figure}
\centering
\includegraphics[width=1.5in]{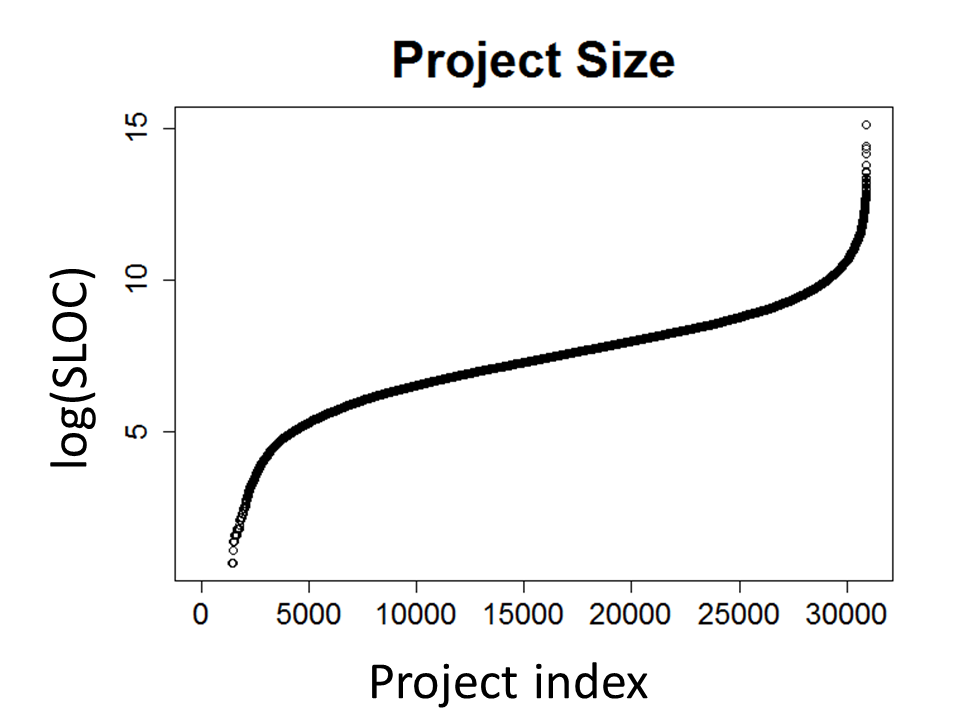}
\includegraphics[width=1.5in]{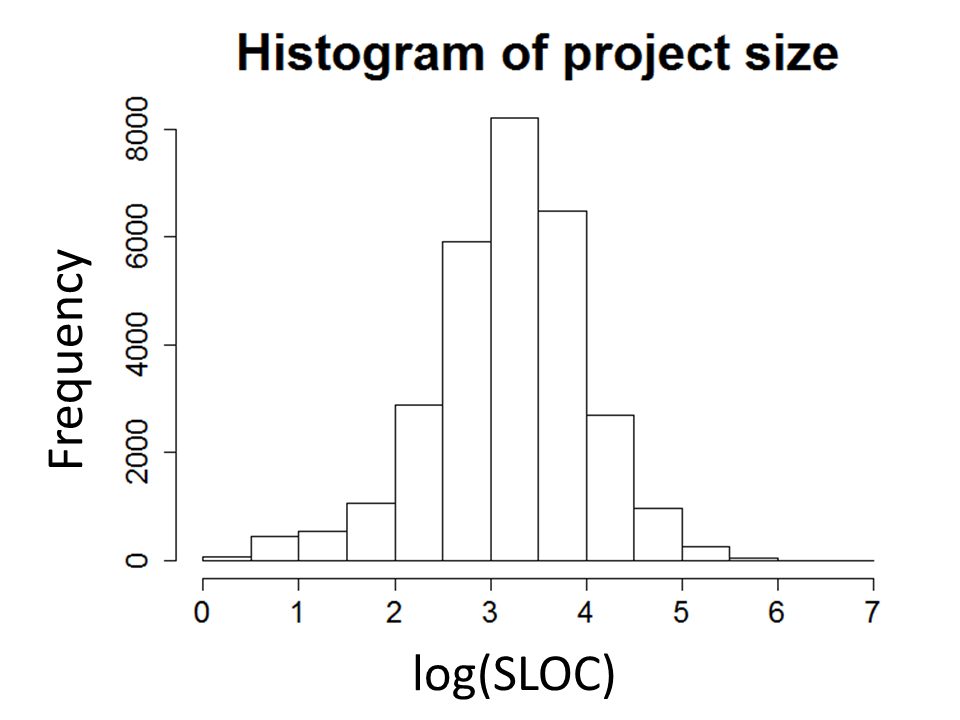}
\caption{Left: Size of the projects in the Google Code dataset, in Source Lines
  of Code (SLOC) when projects are ordered by increasing size. Right:
  Histogram of the size of the projects, in log scale.}
\label{fig:projects-size}
\end{figure}

Although the complete dataset contains projects from Google Code,
Sourceforge and Apache, for this study, we restricted the analysis to
the projects from Google Code only. The main properties of the Google
Code dataset are presented in
Table~\ref{tab:dataset}. Figure~\ref{fig:projects-size} shows the size
of the projects, from smallest to largest, as well as the histogram of
project sizes in the dataset.

\begin{table}
\centering
\caption{Main metrics of the Google Code dataset. }
\begin{tabular}{|l|r|}\hline
            & Google Code \\ \hline
Projects    & 30,914      \\ \hline
Classes     & 3,060,853   \\ \hline
Interfaces  & 274,745     \\ \hline
Methods     & 19,358,490  \\ \hline
SLOC        & 221,194,474 \\ \hline
Median SLOC & 1,570       \\ \hline
\end{tabular}
\label{tab:dataset}
\end{table}

This study's granularity is a ``project.'' For the purposes of this
study, a project is the collection of Java source code files that were
found in each Google Code Project Hosting's project pages. For
example, the project named {\texttt 1cproject} was hosted at
https://code.google.com/p/1cproject/, and its source code was
available at
\\ https://code.google.com/p/1cproject/source/browse/\\ 
The ``project,'' in this case, consists of all Java source files found
under source control in {\em trunk}. When the project included jar
files, those were considered potential dependencies, not part of the
project itself.\footnote{This paragraph is written in the past tense,
  because Google Code is slated to become unavailable soon.}

\vspace*{0.3cm}
\noindent
{\bf Availability of Data and Tools}

\noindent
The Sourcerer infrastructure and tools are available from
Github~\cite{Sourcerer:2015}, and have been described before in our
prior papers~\cite{Ossher:2009,Bajracharya:2014}. Besides those two
prior publications, a publicly available tutorial explains the
processing pipeline of the Sourcerer tools with concrete
examples~\cite{Lopes+Ossher:2012}. Additionally, the artifact
associated with this paper contains all the Sourcerer tools and a
small sample repository of projects, meant to illustrate the
processing pipeline by which raw source code is converted into a
relational database for static analysis, such as that in this
paper. Note that only a small repository is included, because the full
repository is 433Gb; its processing into a relational database took
approximately 3 weeks of computing on a 24-core, 128 Gb RAM server.

Researchers wanting to reproduce this study, or wanting to study other
facets of this data, can start by downloading the artifact associated
with this paper, and running the Sourcerer tools installed in it on
the included sample repository; then, they can download the full
repository from our Web site~\cite{Lopes+Ossher:2012} and run the
Sourcerer tools on it.

Having done all this processing ourselves, we are making the processed
datasets available to other researchers. The several representations
of the Sourcerer 2011 dataset, including the full repository and the
database, are publicly available for download
from our Web site~\cite{Lopes+Ossher:2012}. Note that this dataset is immutable; it
was collected once in 2011, and we do not plan to collect later
versions of the projects. The CSV file upon which statistical analysis
of this study was done is included in the artifact.

\section{Statistical Analysis Methods}

This section explains the main statistical methods that were used in
this study.

\subsection{Linear vs. Log Scales}

\begin{figure}
\centering
\includegraphics[width=1.5in]{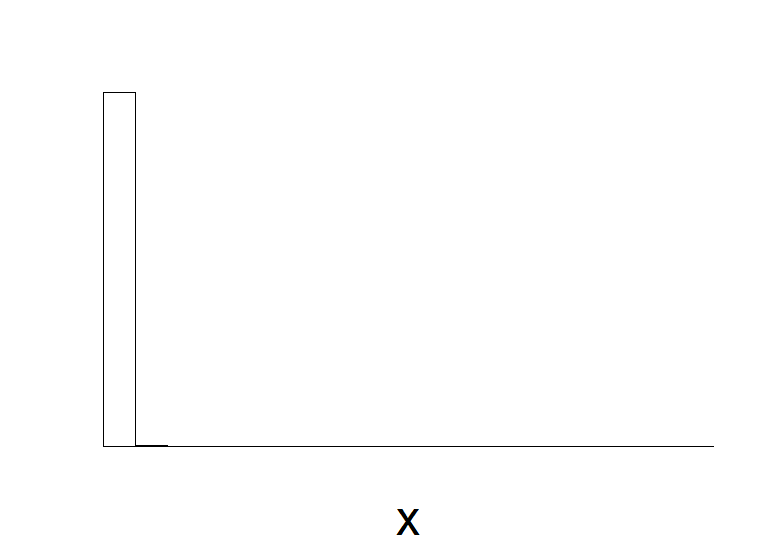}
\includegraphics[width=1.5in]{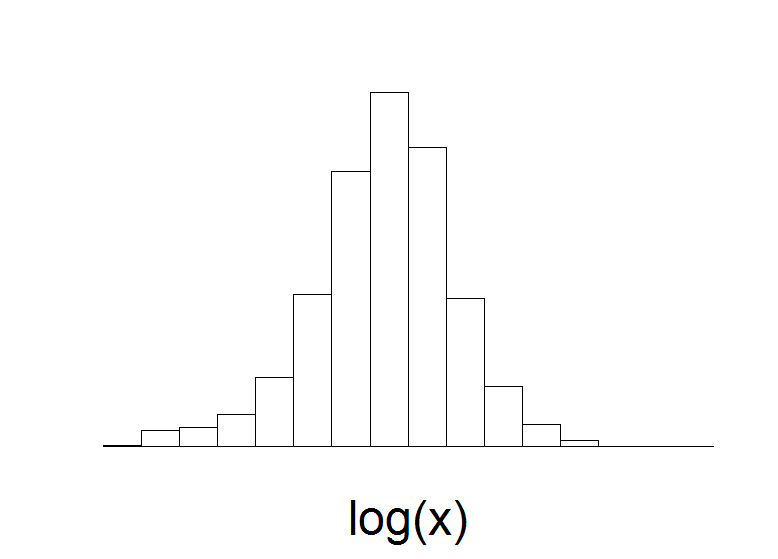}
\caption{Histograms of log-normal data when plotted in {\em linear}
  scale (left) and {\em log} scale (right).}
\label{fig:lognormal}
\end{figure}

As mentioned in Section \ref{sec:relwork}, when dealing with large
ecosystems of software artifacts, the data is expected to be highly
skewed in almost every dimension. That is also the case in the Google
Code data. Figure~\ref{fig:lognormal} shows a generic illustration of
skewness in the data: the left histogram shows that the vast majority
of data points have small values of $X$, where $X$ is some measured
feature of the dataset; in transforming the data into log scale,
however, we can see an almost perfect log-normal distribution (right
histogram). When this holds, it would be ill-suited to use normal
statistics in linear space, but we can proceed to apply normal
statistics in log space. This is a critical step in analyzing
these ecosystems.

\begin{figure}
\centering
\includegraphics[width=1.5in]{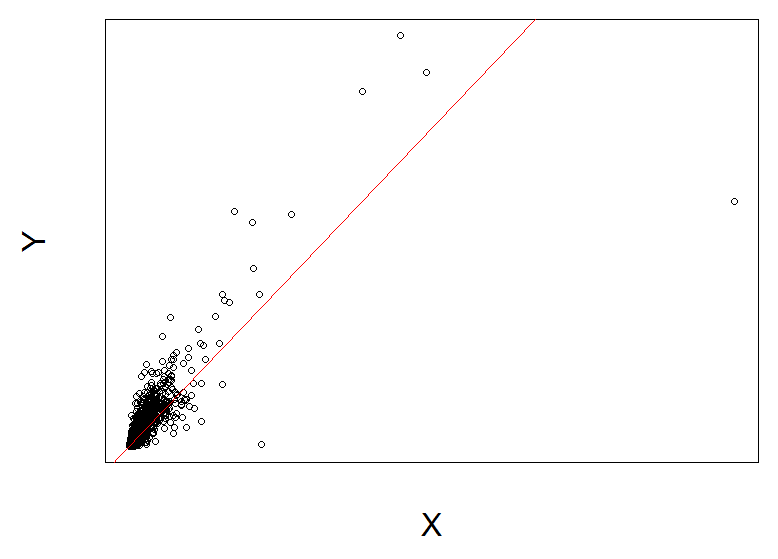}
\includegraphics[width=1.5in]{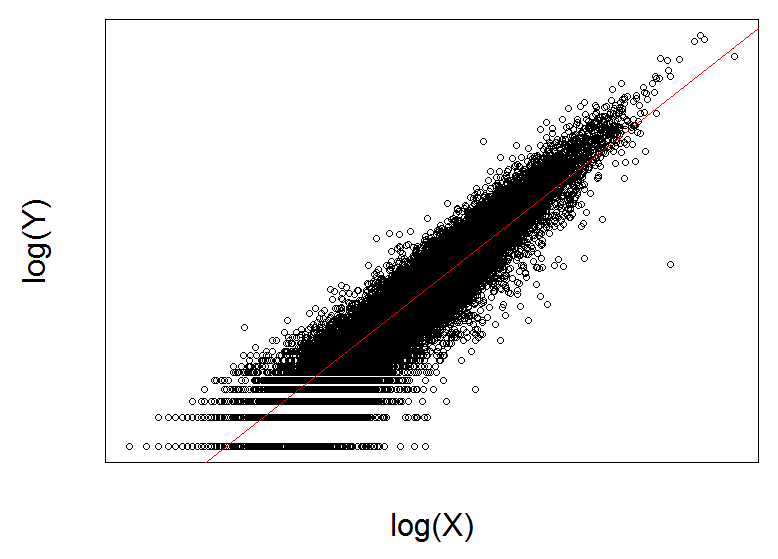}
\caption{Scatterplots of Y against X when both X and Y are
  log-normal data. On the left: plot in {\em linear} scale of
  X and Y; on the right: plot in {\em log} scale of X and Y.}
\label{fig:lm}
\end{figure}

\subsection{Linear Regression Models}

The main statistical tool we use in this study is the linear
regression model. Linear regression tries to find the best linear
model (i.e. a line) that fits the data. Figure~\ref{fig:lm}
illustrates our use of this statistical tool. On the left, we see a
scatterplot of some feature $X$ against some other feature $Y$ plotted
in linear scale of both X and Y. The plot also shows the best fit line
resulting from linear regression of the data. On the right, we see the
scatterplot of the same features $X$ and $Y$ but plotted in log scale,
along with the best fit line. In both cases, the line is given as
$y\_values = \alpha + \beta x\_values$. However, the plot on the right
being in log scale, the straight line represents $log(y) = \alpha + \beta
log(x)$. Transforming this back to linear space gives the following
non-linear (exponential) relation between $X$ and $Y$:

\begin{equation} 
\label{eq1}
y=e^{\alpha} x^{\beta}
\end{equation}

When the relation between two features is {\em non-linear}
and, specifically, exponential, some observations are at hand:
\begin{itemize}
\item When $\beta=1$, the relation between $X$ and $Y$ degenerates to
  linear.
\item Any value of $\beta \neq 1$ indicates an exponential relation
  between the two features. Small variations in $\beta$ represent
  large variations of $Y$ against $X$ in linear space.
\item $\beta > 1$ indicates a superlinear relation, i.e. $Y$ grows
  exponentially faster as $X$ grows.
\item $\beta < 1$ indicates a sublinear relation, i.e. $Y$ grows
  exponentially slower as $X$ grows.
\end{itemize}

\subsection{Goodness of Fit}

One critical part of linear regression is the goodness of fit, that
is, how well the line fits the data. $R^{2}$, prononced R-squared, is
a statistic that measures how successful the fit is in explaining the
variation of the data.\footnote{$R^{2} = 1 - \dfrac{SSE}{SST}$, where
  $SSE$ is the residual sum of squares and $SST$ is the total sum of
  squares.} For example, $R^{2}=0.92$ means that the fit explains 92\%
of the total variation in the data. A value of 1 would be the perfect
fit.

\begin{figure}
\centering
\includegraphics[width=3.5in]{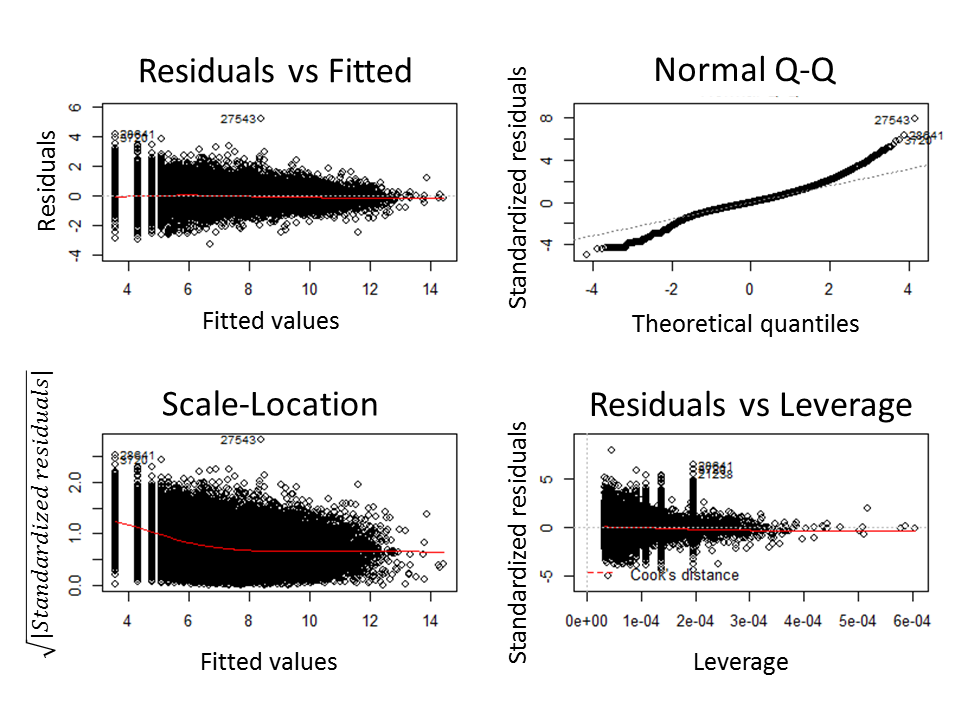}
\caption{Residuals plots. Top-left: Residuals vs Fitted. Top-right:
  Normal QQ. Bottom-left: Scale-location (aka spread). Bottom-right:
  Residuals vs Leverage. }
\label{fig:residuals}
\end{figure}

However, due to how it is calculated, there are several limitations
for what $R^{2}$ can explain. Depending on the characteristics of the
data, $R^{2}$ can have a low prediction value. In order to verify
this, it is important to analyse the {\em residuals} of the linear
regression models. Figure~\ref{fig:residuals} illustrates the kinds of
residuals plots that we analyze to check whether the linear models are
appropriate or not. The {\em Residuals vs Fitted} plot (top-left) is
the most important one. A good fit should result in this plot showing
randomly distributed data around the horizontal line at the origin,
meaning that what's left from the fit is unbiased noise -- this
particular plot shows that. When this doesn't happen, then the linear
model may not be appropriate to explain the data, even if $R^{2}$ is
high. The {\em Normal QQ} plot (top-right) illustrates assumptions
about normality of the residuals in the model. When the dots all fall
in the straight diagonal, then the residuals fit exactly a normal
distribution, which is the ideal case. This particular plot shows a
symmetrical light-tailed normal distribution of the residuals, which
is acceptable. In general, some deviation from the norm is to be
expected, particularly near the ends. The {\em Scale-Location} plot,
also known as {\em spread}, illustrates the variance of the Y variable
along the X variable. A flat line means that the variance is constant
along X, which is the ideal case for linear regression. This particular
plot shows that there is more variance for lower values of X, and then
the variance evens out. This kind of small deviation from the ideal is
acceptable. Finally, {\em Residuals vs Leverage} illustrates the {\em
  leverage} (influence) that the data points had on the fitness
process. This plot serves to identify potential outliers that may have
had undue influence in the model. We want the points to fall as close
as possible to the horizontal line at origin, and not to fall outside
Cook's distance. That is the case with this particular plot.

\subsection{Binned Analysis}

When the residuals of the linear models show potential problems with
the model, that means that the simple linear regression models are
missing important characteristics of the data. In those cases, we try
to perform binned analysis instead of analysis on the whole data. This
analysis is meaningful when the data in the bins shows normal
distributions. When that is the case, we compare the differences of
means among the bins using Welch two sample t-test on a 95\%
confidence interval in order to extract more meaningful insights.

\section{Findings}
\label{sec:results}

This section presents the main findings of our study. It starts with
observations regarding the size of the modules, then their complexity,
the use of inheritance, and finally the kinds of dependencies the
modules have. It should be noted that all linear models and
correlations presented here are statistically significant, with
p-values $<<$ 0.0001.

\subsection{Module Size}
\label{sec:modulesize}

RQ1: {\em Are modules of larger systems larger, or smaller, than modules
  of smaller systems, or are there no statistically significant
  differences?}

In Java, the modular, replaceable units are classes and interfaces, so
we use the word {\em module} to mean either a class or an
interface. The bivariate analyses used to study the above question
are: SLOC vs. Modules, Methods vs. Classes and Constructors
vs. Classes. We know from several previous studies that these pairs of
metrics are strongly positively correlated. But what exactly is the nature
of these relations?

\begin{figure}
\centering
\includegraphics[width=3in]{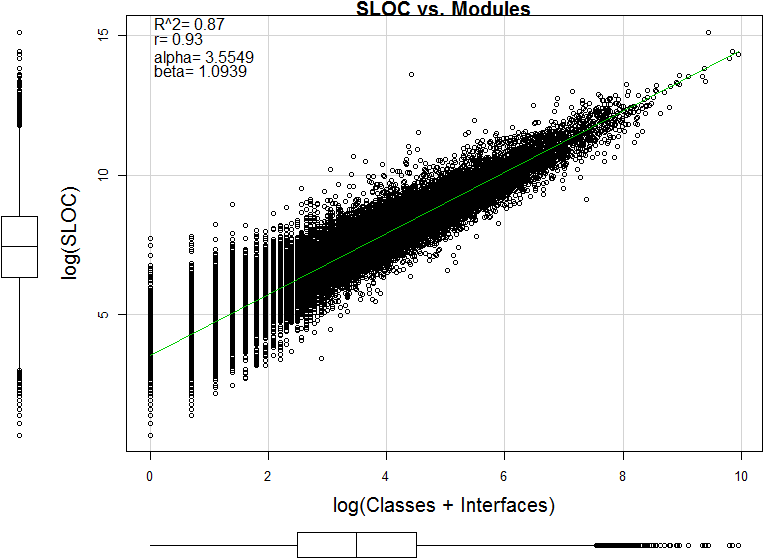} \\
\vspace*{0.5cm}
\includegraphics[width=3in]{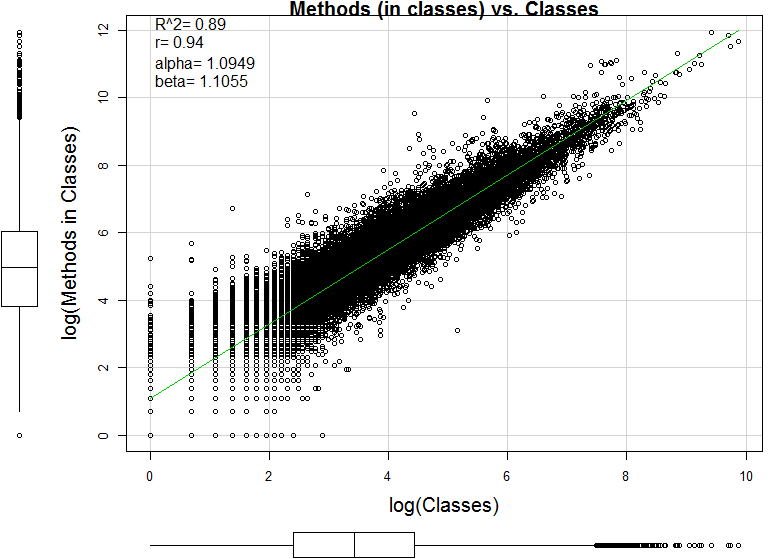} \\
\vspace*{0.5cm}
\includegraphics[width=3in]{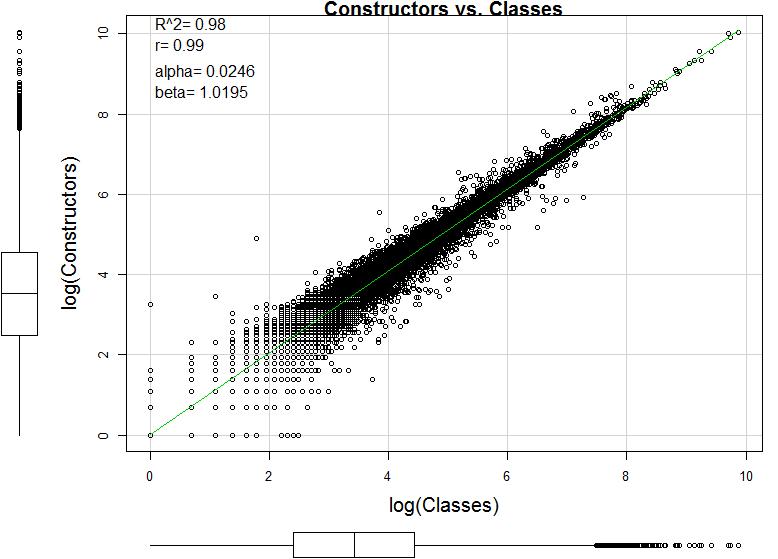}
\caption{Top: SLOC vs. number of modules. Middle: Number of methods
  vs. number of classes. Bottom: Number of constructors vs. number
  of classes.}
\label{fig:modulesize}
\end{figure}

Figure~\ref{fig:modulesize} sheds some light into this question. On the
top, the size of each project in SLOC is plotted against its number of
classes and interfaces. As expected there is a very strong linear
correlation ($r=0.93$). Moreover, with 87\% linear fitness (in log
space), it appears that as the number of modules grows, the lines of
source code also grows but at exponentially higher pace. Specifically,
using equation~\ref{eq1},

\begin{center}
$SLOC=e^{3.5549} Modules^{1.0939}$
\end{center}

For example, a project with 10 modules is predicted to have close to
434 SLOC; a project with 100 modules is predicted to have not just
4,340 but close to 5,391 SLOC, so considerably more than 10 times
what's expected of a project with 10 modules; a project with 1,000
modules is predicted to have close to 66,923 SLOC, again considerably
more than 10 times what's expected of a project with 100 modules;
etc. The growth of SLOC is exponential with number of modules, and
even though the exponent (1.0939) is close to 1, the small 0.0939
difference results in large differences in the linear space.

Where do all these extra lines of code go? The middle plot in
Figure~\ref{fig:modulesize} explains it. The plot shows the number of
methods declared in classes vs. the number of classes. Again, as
expected, there is a strong positive linear correlation
($r=0.94$). Moreover, with 89\% linear fitness, it appears that as the
number of classes grows, the number of methods grows
exponentially. Specifically, 

\begin{center}
$Methods=e^{1.0949} Classes^{1.1055}$
\end{center}

For example, a project with 10 classes is predicted to have close to
38 methods; a project with 100 classes is predicted to have not just
380 but close to 486 methods; a project with 1,000 classes is
predicted to have close to 6,195 methods; etc. Again, the growth of
the number of methods is exponential, not linear. 

A similar exponential growth can be observed for constructors
vs. classes (Figure~\ref{fig:modulesize}, bottom). The relation in that
case is $Constructors = e^{0.0246} Classes^{1.0195}$. In this case,
$\beta$ is very close to 1, so this is almost a linear function.

In short: projects with {\bf more modules} have disproportionately more
lines of code than projects with less modules, which means that they
have {\bf larger modules}. Moreover, the extra lines of code seem to
be grouped in disproportionately {\bf more methods} and, to a lesser
degree, constructors, {\bf per class}.

Table~\ref{tab:modulesize} summarizes the statistical principles
inferred from the data related to the effect of size. $\alpha$ and
$\beta$ are the coefficients for equation~\ref{eq1}; $r$ is the
Pearson correlation coefficient of the data in log scale; $R^{2}$ is
the fitness of the line in log space. The residuals of the linear
model can be found in Figures \ref{fig:res_SLOC_Modules},
\ref{fig:res_Methods_Classes} and
\ref{fig:res_Constructors_Classes}. All of them show good strength of
the model.

\begin{table}
\centering
\caption{Analysis of project size.}
{\small
\begin{tabular}{|l|r|r|r|r|l|}\hline
Analysis            & $\alpha$ & $\beta$& $r$  & $R^{2}$ & Space\\ \hline
SLOC vs. Modules    & 3.5549   & 1.0939 & 0.93 & 0.87 & log-log \\ \hline
Meths. vs. Classes  & 1.0949   & 1.1055 & 0.94 & 0.89 & log-log \\ \hline
Constrs. vs. Classes & 0.0246  & 1.0195 & 0.99 & 0.98 & log-log \\ \hline
\end{tabular}
}
\label{tab:modulesize}
\end{table}

These observations explain apparent inconsistencies in the literature
over the past few years. As described in Section~\ref{sec:relwork},
different studies of Java corpora have reported different average
methods per class. This could potentially be explained by our
findings: a corpus that is dominated by smaller projects will have
lower average methods per class than a corpus dominated by larger
projects.

\subsection{Module Types}
\label{sec:moduletype}

\begin{figure}
\centering
\includegraphics[width=3in]{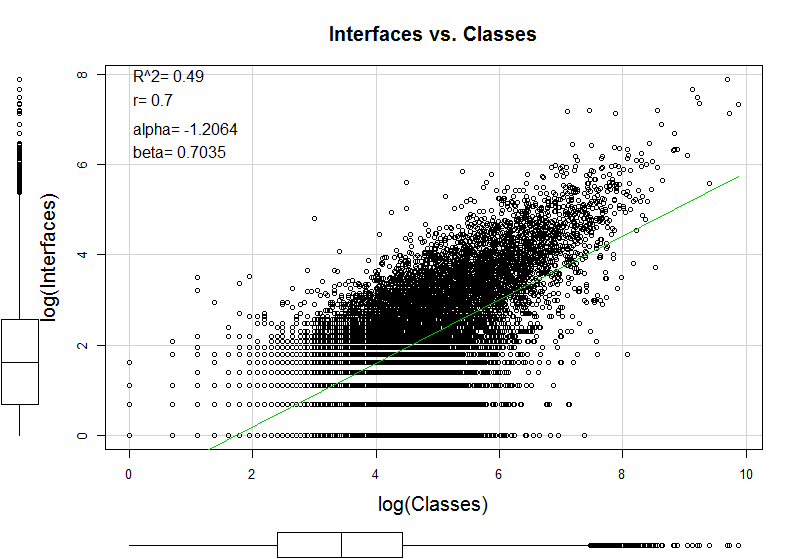}
\caption{Interfaces vs. classes in log scale.}
\label{fig:moduletype}
\end{figure}

RQ2: {\em Is there a statistically significant variation in the mix of
  classes and interfaces for projects of different scales?}

The answer to the question involves dealing with data that shows high
variance. We started by regressing the number of interfaces against
the number of classes in each project, similarly to what we did for
the previous question. Figure~\ref{fig:moduletype} shows
the non-linear model:

{\em Interfaces} $= e^{-1.2064}Classes^{0.7035}$

$R^2 =0.49$ is not too good of a fit. Visual inspection of the plot
and the fitted straight line exposes weaknesses of this simple linear
model, particularly at the edges of the data: for projects with very
small and very large classes, the model underestimates the number of
interfaces. Clearly, the relation between the number of classes and the
number of interfaces in this ecosystem is not properly explained by
a simple exponential function.

The observations from visual inspection are confirmed in the plots of
the residuals in Figure~\ref{fig:res_Interfaces_Classes} (Appendix
B). The {\em Residuals vs Fitted} plot, in particular shows a bend in
the residual data, rather than a straight line. This is indicative
that the linear model is missing a non-linear component. The shape of
the bend suggests a parabola, so we add an additional transformation
of the X variable, specifically $log(X)^2$, and perform a linear
regression on that transformed space ($log(y) \sim log(x)^2$). This
additional transformation introduces non-monotonicity (see Appendix 
Figure~\ref{fig:log_squared}). Transforming this back to linear
space, we are establishing the relation:

\begin{equation} 
\label{eq2}
y=e^{\alpha} x^{\beta log(x)}
\end{equation}

\begin{figure}
\centering
\includegraphics[width=3in]{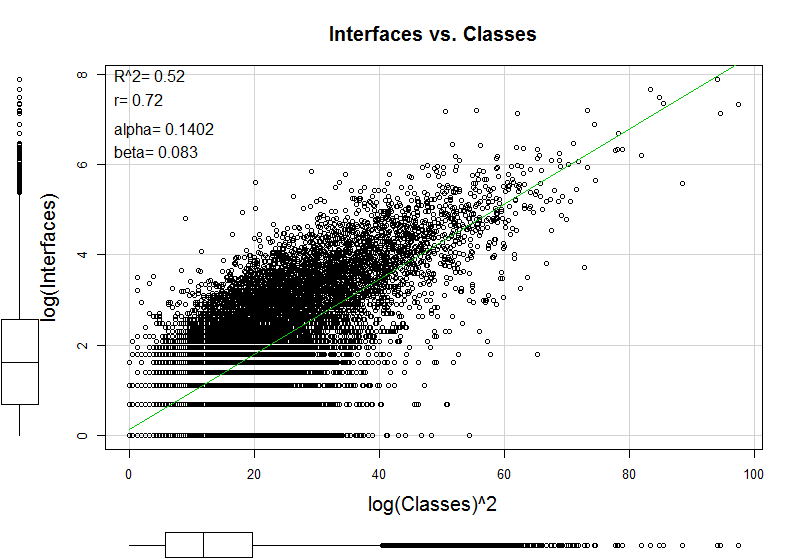}
\caption{Interfaces vs. classes in log$^2$ scale.}
\label{fig:moduletype2}
\end{figure}

The plot is shown in Figure~\ref{fig:moduletype2} and the residuals
plots are in Appendix B, Figure~\ref{fig:res_Interfaces_Classes2}. As
can be seen, this is a better fit, with $R^2=0.52$. The bias in the
fit that transpired with the bend in the residuals plot is now
practically eliminated. Given the parameters, this model predicts 

{\em Interfaces} $= e^{0.14} Classes^{0.083 log(Classes)}$

Note that due to the non-monotonicity illustrated in
Figure~\ref{fig:log_squared}, this model establishes a variable ratio
between classes and interfaces depending on project
size. Specifically, smaller projects have a much higher ratio\\
$Interfaces/Classes$. \\
For example, for a project with 10 classes, the
model predicts 1.79 interfaces ($\sim$ 18\% ratio); 50 classes $\succ$
4.1 interfaces ($\sim$ 8.2\%); 100 classes $\succ$ 6.69 interfaces
($\sim$ 7\%); 1000 classes $\succ$ 60.4 interfaces ($\sim$
6\%). Acording to the model, the ratio proceeds to increase again for
very large projects. For example 10,000 classes $\succ$ 1,314
interfaces ($\sim$ 13\%). Our dataset includes only 6 projects that
contain over 10,000 classes each, so the model may not be precise for
this end of the size spectrum.

Another way of analyzing this data is to make a binned analysis.  For
that, we divide the data into 5 bins on the number of classes: very
large, large, medium, small and very small. We then compute the ratio
$Interfaces/Classes$ for all the projects, and compute the means of
the ratios in each bin. The results are shown in
Table~\ref{tab:bin_moduletype}.

\begin{table}
\centering
\caption{Bins for analysis of {\em Interfaces/Classes}. Mean
  and SD values are in log scale.}  {\small
\begin{tabular}{|l|l|r|r|r|}\hline
  Bin    & \# Classes     & Projects & Mean (linear\%) & SD   \\ \hline
V. Large & $>$ 5,000      & 17    & -2.47 (8.5) & 0.87 \\ \hline
Large    & 1,000 -- 5,000 & 419   & -2.83 (5.9) & 1.00 \\ \hline
Medium   & 100 -- 1,000   & 5,762 & -2.77 (6.3) & 1.05 \\ \hline
Small    & 20 -- 100      & 11,715& -2.49 (8.3) & 0.92 \\ \hline
V. Small & $<$ 20         & 11,557& -1.68 (18.6)& 0.78 \\ \hline
\end{tabular}
}
\label{tab:bin_moduletype}
\end{table}

Finally, we perform a Welch two sample t-test on the differences of
means to check whether the differences exist and are statistically
significant. The tests show statistical significance (p $<<$ 0.0001)
on the mean differences between medium and small, and between small
and very small. The other differences are not statistically
significant at 95\% confidance level.

The numerical values of the binned analysis are consistent with those
from the linear regression model that places the number of interfaces
as a continuous function of the number of classes as by
equation~\ref{eq2}. This adds strength to the result.

One possible explanation for why smaller projects have
disproportionately more interfaces is that they have an investment in
modeling entities with interfaces without having enough
implementations of those entities to pay off the investment. In
medium-to-large projects, that investment pays off, as more classes
provide alternative implementations of the interfaces. The higher
ratio in very large projects is not statistically significant, so no
conclusions should be made on whether that holds in general or only in
this particular set of 17 very large projects.

\subsection{Internal Complexity}
\label{sec:modulecomplexity}

RQ3: {\em Do larger projects have more method calls or use more unsafe
  operations than smaller projects, or are there no statistical
  significant differences?}

A recent study by Landman et al.~\cite{Landman2014} showed that
there seems to be no correlation between the size of Java projects
(measured in SLOC) and the cyclomatic complexity of their methods. We
go one step further to investigate other potential sources of
complexity in code: the number of outgoing methods calls, the
number of {\texttt instanceof} statements and the number of unsafe
type casts in each project. 

\begin{table}
\centering
\caption{Analysis of code complexity}
{\small
\begin{tabular}{|l|r|r|r|r|l|}\hline
Analysis           & $\alpha$ & $\beta$& $r$  & $R^{2}$ & Space\\ \hline
Calls vs. Methods  & 1.64  & 1.00 & 0.94 & 0.89 & log-log\\ \hline
Inst.of vs. Methods& -2.77 & 0.84 & 0.70 & 0.49 & log-log\\ 
                   & -0.41 & 0.01 & 0.72 & 0.52 & log-log$^{2}$\\ \hline
Casts. vs. Methods & -1.81 & 1.00 & 0.83 & 0.68 & log-log \\ 
                   & -0.49 & 0.36 & 0.84 & 0.70 & log-log$^{1.4}$ \\ \hline
\end{tabular}
}
\label{tab:modulecomplexity}
\end{table}

Like for module types, some of the data here also has a considerable
variation. Table \ref{tab:modulecomplexity} summarizes the results.
We found no evidence that larger projects use more unsafe features of
Java than smaller projects, and we found some weak evidence that the
contrary may happen.

\subsubsection{Method Calls vs. Methods declared in classes}
\label{sec:callsvsmethods}

In the case of method calls, there is a fairly strong fit of the
linear model ($R^{2}=0.89$), and the residuals plots show no warning
signs (Appendix B Figure\ref{fig:res_Calls_Methods}). The exponent
$\beta=0.9971$, however, is very close to 1, which means that the
relation is essentially linear at a rate of $e^{1.64}=5.1$ calls per
method. For example, according to the model, a project with 50 methods
has 255 method calls; a project with 500 methods has 2,531 method
calls; a project with 5,000 methods has 25,144 method calls. 

\subsubsection{Instanceof statements vs. Methods declared in classes}
\label{sec:instanceofvsmethods}

In the case of {\texttt instanceof}, the linear model in log-log space
is not that good ($R^{2}=0.49$), and the residuals plots show some
warning signs (Appendix B
Figure~\ref{fig:res_Instanceof_Methods}). Similarly to what was done
for the previous analysis, we transformed the X axis (methods) with an
additional square function, and the fit improved to $R^{2}=0.52$
(residuals in Appendix B
Figure~\ref{fig:res_Instanceof_Methods2}). This yields the
relation\\ 
{\em Instanceof} $= e^{-0.14} Methods^{0.0702 log(Methods)}$

Again, this function is not monotonic, and therefore results in a
non-monotonic average number of {\texttt instanceof} statements per
method, depending on the total number of methods of the projects: the
ratio starts high for projects with just a few methods (e.g. 0.21 for
projects with 5 methods) and decreases sharply for projects with very
small number of methods ($< 20$); it then continues to decrease but
more gently, reaching a minimum of 0.025 {\texttt instanceof}
statements per method for projects with around 1,000 methods
(i.e. almost 10 times less than for projects with 5 methods); from
then on, it increases again, but slowly. Its predicted value is 0.06
{\texttt instanceof} statements per method for projects with 50,000
methods (of which there are 17 in the dataset).

\subsubsection{Type casts vs. Methods declared in classes}
\label{sec:castsvsmethods}

In the case of casts, the linear log-log model is also not that good
($R^{2}=0.68$, see also Appendix B
Figure~\ref{fig:res_Casts_Methods}). We then tried a few
transformations, and found $log^{1.4}$ to produce very good residuals
plots (Appendix B Figure~\ref{fig:res_Casts_Methods2}) and a better
$R^{2}=0.70$. This function has a similar behavior as the one
explained for {\texttt instanceof} in terms of monotonicity, but the
minimum (0.13 casts per method) happens a bit earlier, at around 500
methods. This value of casts per method is roughly 10 times less than
the value for projects with 5 methods.

\subsubsection{Discussion}

Combined, and along with the study by Landman et al., these results
show that there is no evidence to support the hypothesis that larger
projects have more complex code. On the contrary, there seems to be a
trend for smaller projects to include proportionally more unsafe
statements of Java. 

The linear model method calls vs. methods (Section
\ref{sec:callsvsmethods}) is a fairly strong fit that shows that the
number of method calls per declared method is roughly constant and
independent of the size of the projects (measured in number of
methods). The other two models (Secitons \ref{sec:callsvsmethods} and
\ref{sec:instanceofvsmethods}) have a less strong fit. That simply
means that their precision as predictors is not too good, but the
trend showing proportionally more unsafe features of Java in small
projects (Section~\ref{sec:castsvsmethods}) is interesting and
statistically significant. We conjecture that this may happen because
developers of non-trivial projects adhere to a stricter discipline of
avoiding these features.

\subsection{Class Composition via Inheritance}
\label{sec:inheritance}

RQ4: {\em Does the scale of the project affect the use of
  inheritance?}

The two linear models used to answer the above question are classes
defined using inheritance (DUI) vs. total classes, and classes that
are inherited from within the project (IF) vs. total classes of each
project. The results are shown in Table~\ref{tab:inheritance} (plots
in Appendix \ref{fig:res_DUI_Classes},
\ref{fig:res_DUI_Classes2} \ref{fig:res_IF_Classes} and
\ref{fig:res_IF_Classes2}.

\begin{table}
\centering
\caption{Analysis of inheritance}
{\small
\begin{tabular}{|l|r|r|r|r|l|}\hline
Analysis         & $\alpha$ & $\beta$& $r$  & $R^{2}$ & Space \\ \hline
{\footnotesize DUI vs. Classes}  & -1.0505  & 1.0159 & 0.92 & 0.85 & log-log \\
                 & -0.5364  & 0.6626 & 0.93 & 0.86 & log-log$^{1.2}$ \\ \hline
{\footnotesize IF vs. Classes}   & -1.9908  & 0.8037 & 0.78 & 0.61 & log-log \\ 
                 & -0.3414  & 0.0903 & 0.80 & 0.64 & log-log$^{2}$ \\ \hline
\end{tabular}
}
\label{tab:inheritance}
\end{table}

As in previous analysis, the residuals plots of the initial linear
regression models showed some warning signs that the models might not
be the best (Appendix B Figures~\ref{fig:res_DUI_Classes2} and
\ref{fig:res_IF_Classes2}). As such, we compensated for the bend in
the residual data by adding an additional non-linear components to the
X axis (classes). 

\subsubsection{Classes Defined Using Inheritance (DUI)}

In the case of DUI classes vs. classes, the better
model is\\ 
{\em DUI} $= e^{-0.5364+0.6626 log(Classes)^{1.2}}$

Also here, the curve of the ratio starts high, decreases sharply, then
decreases slowly up to a minimum, then increases again. In the case of
these parameters, the minimum is around 10 classes, with 35\% of them
DUI. According to this model, projects with 2 classes have on average
0.9 of them defined using inheritance (45\%); 10 classes $\succ$
3.5 (35\%); 100 classes $\succ$ 37 (37\%); 1,000 classes $\succ$ 493
(49\%); 5,000 classes $\succ$ 3,379 (68\%); etc.

A project with 100 classes, 65\% of them DUI, is far from the norm,
but if the number of classes is close to 5,000, then that percentage
of DUI is close to the norm.

\subsubsection{Classes Inherited From (IF)}

In the case of IF classes, the better model is\\ 
{\em IF} $= e^{-0.3414} Classes^{0.0903}$

In the case of these parameters, the minimum is around 100 classes,
with 5\% of them DUI. According to this model, projects with 10
classes have on average 1.1 inherited from (11\%); 100 classes
$\succ$ 4.8 (5\%); 1,000 classes $\succ$ 52.8 (5\%); 5,000 classes
$\succ$ 3,379 (10\%); etc.

\subsection{Dependencies}
\label{sec:dependencies}

RQ5: {\em Do larger projects use disproportionately more, or fewer,
  modules than smaller projects? How does project efferent coupling
  vary with size? Are there statistically significant differences in how
  types from JDK/internal/external libraries are used in projects of
  varying sizes?}

To answer these questions, we first look at the growth of the number of
distinct types used by projects vs. the growth of the number of
modules\footnote{Again, we use the term modules = types = classes +
  interfaces.} declared in the projects. In counting the number of
modules used, we count the number of {\em distinct} modules, so
modules used multiple times in a project are counted only once. We
then look deeper into the origins of the used modules.

\subsubsection{Used Modules vs. Declared Modules}

Projects use a variety of modules, some of them declared internally,
others provided by the JDK and others provided by external
libraries. Again, we know that the number of modules used in a project
is highly correlated with the size of the project. We are interested
in studying the underlying trend function, and whether it is linear or
super-/sub-linear. Project size here is measured by
the number of declared modules in it.

In analyzing the initial linear regression model in log-log space, it
was visible that it suffered from a small bend in the residuals (see
Appendix B Figure~\ref{fig:res_Used_Modules}.) We then compensated
for it by applying a polynomial of size $1.2$, which eliminated the
bend, producing a better model. Table~\ref{tab:moduleuse} summarizes the
parameters.

\begin{table}
\centering
\caption{Analysis of used modules}
{\small
\begin{tabular}{|l|r|r|r|r|l|}\hline
Analysis          & $\alpha$ & $\beta$& $r$  & $R^{2}$ & Space \\ \hline
{\footnotesize Used vs. Declared} & 2.006   & 0.7357 & 0.93 & 0.87 & log-log \\
                  & 2.335   & 0.4863 & 0.93 & 0.87 & log-log$^{1.2}$ \\ \hline
\end{tabular}
}
\label{tab:moduleuse}
\end{table}

The better model establishes the following relation between the number
of used modules and the number of declared modules in a project:

{\em Used} $= e^{2.3353 + 0.4863 log(Modules)^{1.2}}$

These parameters define a sub-linear relation between the number of
used modules and the number of declared modules in a project, meaning
that the number of distinct used modules increase disproportionately less than
the number of declared modules in projects. According to this model, 
projects with 10 declared modules use on average 39 modules; 100
declared $\succ$ 216 used; 1,000 declared $\succ$ 1,450 used; etc. The
number of used modules grows slower than the number of declared modules.

This result is intriguing, as it was unclear what to expect. The
result makes sense when the addition of a dependency (external or
internal) is correlated with the addition of multiple modules internal
to the project; the causal relation is unclear, and there may be
unknown confounding factors behind this correlation. 

Theoretically, according to this model, there is a scale point at
which the number of used modules is less than the number of declared
modules, which means that some declared modules would not be used,
just declared. That point is around 50,000 declared modules. Our
dataset does not contain any project that large, but we found 753
projects where the number of declared modules is larger than the
number of used modules, so this situation is not rare. An analysis of
this set of projects shows that they are statistically larger than the
average of the whole dataset, and that it contains a disproportionate
number of very large projects -- 17 out of the 59 projects with more
than 3,000 declared modules are in this subset of projects that have
higher number of declared modules than used modules. It is possible
that these cases correspond to utility frameworks.

However, the opposite result, if it had been observed, could also be
explained. That is, one could imagine that the number of used modules
would grow faster than the number of declared modules. In this case,
the addition of a dependency (external or internal) would not be
correlated with additional internal modules, and, instead, it would
simply correlate with the addition of methods in existing modules that
use that new entity. That is not the case in this ecosystem: more
methods seem to exist for defining additional functionality with
existing dependencies than new methods are added for using additional
dependencies. (plots omitted for space reasons)

\subsubsection{Efferent Coupling vs. SLOC}
\label{sec:coupling}

The efferent coupling of an entire project is given by the number of
external modules (classes+interfaces) that the project uses. Here we
study its exact relation with project size given in SLOC. This
analysis targets the well-known correlation between efferent coupling
metrics and size of artifacts, in general. Table~\ref{tab:coupling}
summarizes the parameters. (Plots are in
Appendix B Figure~\ref{fig:res_Coupling_SLOC})

\begin{table}
\centering
\caption{Analysis of efferent coupling of projects}
{\small
\begin{tabular}{|l|r|r|r|r|l|}\hline
Analysis          & $\alpha$ & $\beta$& $r$  & $R^{2}$ & Space \\ \hline
{\footnotesize Coupling vs. SLOC} & 0.1176   & 0.5641 & 0.91 & 0.82 & log-log \\ \hline
\end{tabular}
}
\label{tab:coupling}
\end{table}

According to this model, the relation is sublinear, i.e. efferent
coupling grows disproportionately slower than SLOC. Also, here,
``normality'' changes with scale, it's not a simple constant ratio.

\subsubsection{Provenance of Used Modules}

In order to find out whether there are differences in the origin of
dependencies among projects of different sizes, we then looked at the
provenance of all classes and interfaces (i.e. modules) that are used
in each project, and regressed them against size of the project, given
by number of declared modules. Table~\ref{tab:provenance} summarizes
the parameters. (All residuals plots can be found in Appendix B,
Figures~\ref{fig:res_Internal_Modules}, \ref{fig:res_JDK_Modules} and
\ref{fig:res_External_Modules})

\begin{table}
\centering
\caption{Analysis of origin of dependencies (I)}
{\small
\begin{tabular}{|l|r|r|r|r|l|}\hline
Analysis          & $\alpha$ & $\beta$& $r$  & $R^{2}$ & Space \\ \hline
{\footnotesize Inter. vs. Modules}& -0.5040 & 1.0037 & 0.96 & 0.92 & log-log \\ \hline
{\footnotesize JDK vs. Modules}   & 1.7405  & 0.5306 & 0.81 & 0.66 & log-log \\ \hline
{\footnotesize Exter. vs. Modules}& 0.7168  & 0.7489 & 0.80 & 0.65 & log-log \\ \hline
\end{tabular}
}
\label{tab:provenance}
\end{table}

Indeed, these parameters show that there are differences. As $\beta$
indicates, larger projects use disproportionately less modules from
external sources ($\beta=0.7489$) and even less from the JDK
($\beta=0.5306$) than smaller projects. They use slightly
disproportionately more internally-defined modules ($\beta=1.0037$)
than smaller projects. 

These numbers are highly driven by the previous result -- in general,
the number of used modules grows slower than the number of declared
modules. That blurs the true ratios of the origin of dependencies as
projects grow, so let us analyze the data in a different way. We can
take module use as the independent variable and module origin as the
dependent variable. This helps us quantify the mix of dependency
provenance as a function of project size given by the number of total
used modules (i.e. a slightly different size metric that is highly
correlated with the number of declared modules). The results are shown
in Table~\ref{tab:dependencies}.

\begin{table}
\centering
\caption{Analysis of origin of dependencies (II)}
{\small
\begin{tabular}{|l|r|r|r|r|l|}\hline
Analysis        & $\alpha$ & $\beta$& $r$  & $R^{2}$ & Space \\ \hline
{\footnotesize Inter. vs. Total}& -2.882  & 1.282 & 0.93 & 0.87 & log-log \\ 
                & -2.821  & 1.275 & 0.93 & NA   & log-log {\tiny (RLM)} \\ \hline
{\footnotesize JDK vs. Total}   & 0.162   & 0.750 & 0.90 & 0.82 & log-log \\
                & 0.153   & 0.756 & 0.90 & NA   & log-log {\tiny (RLM)} \\ \hline
{\footnotesize Exter. vs. Total}& -1.585  & 1.072 & 0.89 & 0.79 & log-log \\ 
                & -1.454  & 1.059 & 0.89 & NA   & log-log {\tiny (RLM)} \\ \hline
\end{tabular}
}
\label{tab:dependencies}
\end{table}

An inspection of the residuals plots (Appendix B Figures
\ref{fig:res_Internal_Total}, \ref{fig:res_JDK_Total} and
\ref{fig:res_External_Total}) suggested that the simple linear model
may suffer from the effect of outliers, particularly on the use of JDK
entities. As such we performed a {\em robust} linear regression model
(RLM), which excludes outliers.\footnote{Robust linear regression does
  not report $R^{2}$.} The new residuals plots (Appendix B Figures
\ref{fig:res_Internal_Total2}, \ref{fig:res_JDK_Total2} and
\ref{fig:res_External_Total2}) still suffer from some left-skewness,
but the {\em Residuals vs. Fitted} plot shows an improvement. Even with
RLM, the model may not be strong for the edges of the data, i.e. for
extremelly small and extremelly large projects.

The results indicate that, as the number of total used modules grows,
projects use disproportionately much less of the JDK ($\beta=0.7559$)
and much more internal ($\beta=1.2822$) modules. The growth in
external dependencies is also disproportionately larger, but less so
than the use of internal modules ($\beta=1.0589$).

In retrospect, this result makes sense: the reason why projects are
larger is that they define more classes and interfaces; those are
likely to be used internally. For large projects, and given that the
amount of types in the JDK is fixed, the relative importance of the
types from the JDK decreases and the importance of internal types
increases. 

But this result exposes an interesting characteristic of
programming-in-the-large: larger projects use much more of their
internal, and potentially less stable, components. Smaller projects
leverage the JDK.

\section{Sampling Biases}

The linear regression analysis in the previous section was performed
over the entire datastet without excluding any of the projects. The
dataset is heavily right-tailed, with a bias towards small projects,
and with only a few very large projects. Since linear regression
learns the paramters $\alpha$ and $\beta$ from the data, the data that
we use influences the exact values of those parameters. As such, it
could very well be that the non-linear effects that have been reported
in the previous section, given by $\beta \neq 1$, could be an artifact
of the many small projects and the very few very large projects in the
dataset forcing that non-linear behavior in order for the models to
fit the data. If that were to happen, there might be a simpler linear
model with $\beta=1$ that could perfectly explain the data ``in the
middle'' containing only projects above and below certain size
thresholds. In other words, we could give up explaining what happens
for the many very small projects, because the variance in them
is very large, and for very large projects, because there aren't that
many, and focus on finding simple models for projects in between.

We investigated this possibility by constructing alternative models
where the parameters are learned from various subsets that exclude
very small and very large projects. We report the result on only one
of the many bivariate analysis of the previous section, specifically
Methods vs. Classes, which had $\beta=1.1055$
(Section~\ref{sec:modulesize}). Table~\ref{tab:alternativemodels}
summarizes the results. The column Subset in that table denotes the
conditions for project inclusion in the set as a range on the number
of classes in the project. The first row is the baseline model given
in the previous section. All of these models result in good residuals
without any warning signs.

\begin{table}
\centering
\caption{Alternative models for Methods vs. Classes}
{\small
\begin{tabular}{|l|l||r|r|r|r|}\hline
Model & Subset      & \#Projects & $\alpha$ & $\beta$ & $R^{2}$  \\ \hline
1     & Baseline    & 30,914    & 1.095    & 1.106   & 0.89 \\ \hline
2     & [10--3,000] & 22,860    & 1.283    & 1.061   & 0.84 \\ \hline
3     & [20--3,000] & 18,239    & 1.335    & 1.051   & 0.82 \\ \hline
4     & [30--3,000] & 15,030    & 1.347    & 1.049   & 0.81 \\ \hline
5     & [50--1,000] & 10,576    & 1.350    & 1.047   & 0.73 \\ \hline
6     & [100--500]  & 5,167     & 1.232    & 1.068   & 0.52 \\ \hline
7     & [10--100]   & 16,712    & 1.222    & 1.081   & 0.63 \\ \hline
8     & [1,000--3,000] & 386    & 0.168    & 1.218   & 0.36 \\ \hline
\end{tabular}
}
\label{tab:alternativemodels}
\end{table}

As shown in the table, all of the alternative models are still
non-linear, with $\beta \neq 1$, but, with the exception of model 8,
the value is lower than the baseline model. Model 5, which excludes
almost $2/3$ of the dataset and includes the projects in the middle of
the dataset, has the lowest $\beta$. But even in that subset, the
non-linear relation between classes and methods can be
observed. According to the parameters of model 5, a project with 50
classes is predicted to have 232 methods, and a project with 500
classes is predicted to have, not 2,320, but 2,583 methods. Any
concerns that the non-linear relation between methods and classes was
an artifact of sampling bias are put to rest with the results shown in
Table~\ref{tab:alternativemodels}.

Given that there can be an unlimited number of models inferred from any
arbitrary subset of the original dataset, the question arises of which
model to use.  

If the goal is to make predictions based on the models, the accuracy
of each model can be tested on test datasets that aren't part of the
data from which the models are learned.
We exemplify such prediction goals by measuring the Normalized Root
Mean Square Error (NRMSE) of each model on the two extremes of the whole dataset
that have been eliminated from the learning part: the very small
projects ($\#Classes < 10$) and the very large projects ($\#Classes >
3,000$); note in Table~\ref{tab:alternativemodels} that those
projects are not part of any subset. NRMSE is given by

\begin{equation}
\label{eq:MSRE}
\begin{split}
RMSE = \sqrt{\frac{\sum\limits_{t=1}^n{(\text{\^y} - y)^{2}}}{n}}\\
NRMSE = \frac{RMSE}{y_{max} - y_{min}}
\end{split}
\end{equation}

The summary of this accuracy analysis can be seen in
Table~\ref{tab:accuracy}. For comparison, we also show the NRMSE of
each model on the entire dataset. Numbers in bold represent the models
that performed the best.

\begin{table}
\centering
\caption{Accuracy of the models, measured in NRMSE}
{\small
\begin{tabular}{|l||r|r|r|}\hline
Model & V.Small          & V.Large       & All               \\
      & (5,472 projects) & (50 projects) & (30,914 projects) \\ \hline
1     & 0.13467          & 0.1500        & {\em {\bf 0.05110}}  \\ \hline
2     & 0.13109          & 0.1229        & 0.05155 \\ \hline
3     & 0.13083          & 0.1210        & 0.05181 \\ \hline
4     & 0.13081          & 0.1208        & 0.05188 \\ \hline
5     & {\bf 0.13078}    & {\bf 0.1204}  & 0.05189  \\ \hline
6     & 0.13171          & 0.1230        & 0.05136  \\ \hline
7     & 0.13184          & 0.1354        & {\bf 0.05135} \\ \hline
8     & 0.21137          & 0.1530        & 0.05700  \\ \hline
\end{tabular}
}
\label{tab:accuracy}
\end{table}

As expected, the model that performs the best for the entire dataset
is model 1, whose parameters were inferred from that same data. This
case doesn't serve to validate the model, it just confirms what was
expected. Excluding that baseline, the model that performs the second
best on the entire dataset is model 7, which contains many small
projects. The real validation comes only on the performance of the
models on the two test sets containing very small and very large
projects, which weren't contained in the learning data. In both cases,
the model that makes the best predictions is model 5, whose parameters
are inferred from a large portion of small/medium/large size projects.

Given these results, model 5, which learns the parameters ignoring the
edges of the data, should be used instead of the baseline model
1. Similar accuracy analysis should be done for all the other
bivariate analysis. It is likely that the best models are always the
ones that learn the parameters ignoring the projects at the edges,
where there is either more variance or uncertainty. Nevertheless, the
most important take away from this section is that {\bf the
  non-linearities exist in the data, independent of which 
  subsets we choose.}

\section{Implications for Software Metrics}
\label{sec:discussion}

Our study was centered around a very simple question: {\em does the scale of
the software system affect the internal structure of its modules or
are modules scale-invariant?}

For the Java ecosystem, the answer is: yes, the scale of the
system affects several aspects of the internal structure of its
modules, and of the way the modules are put together. Among those, the
number of methods per class, the number of LOCs per module, the use of
inheritance and mix of dependencies stand out. Going back to the LEGO
metaphor, it is as if large Java projects have injected stronger
coupling material and more hooks into the [larger] software bricks.
These findings have profound implications for software research,
especially quantitative studies of software artifacts. We
discuss them here.

As mentioned in Section \ref{sec:confusingeffect}, size has been the
source of much confusion in software studies. As noted several times
in the literature, many software metrics -- for
example, Weighted Methods per Class (WMC) and (efferent and afferent)
Coupling, just to mention two -- are correlated with size, so their
statistical power is very weak when size metrics are
available. We explain how to properly normalize for size with one
example metric: WMC.

\begin{figure}
\centering
\includegraphics[width=3in]{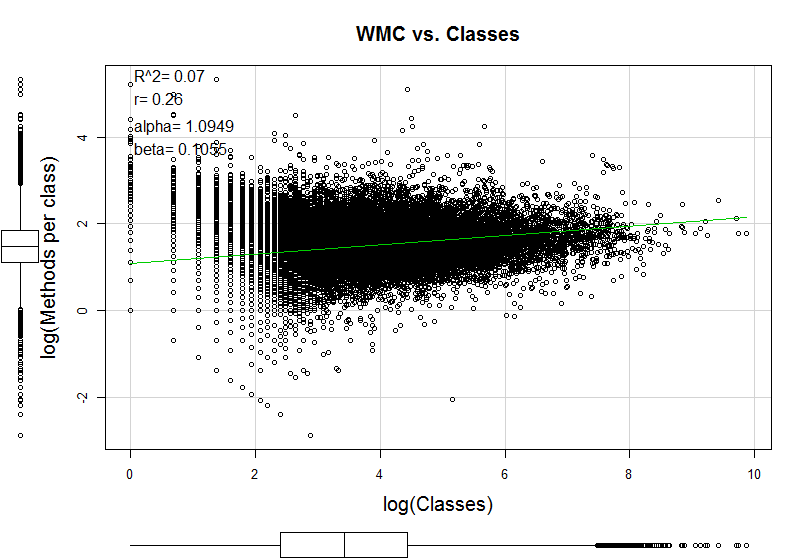}
\caption{Correlation: how WMC grows with the number of classes.}
\label{fig:wmc_classes}
\end{figure}

Figure~\ref{fig:wmc_classes} shows the regression of WMC
vs. Classes in our dataset, a confirmation of what we already know
about the existence of these correlations. In our data, the Person
correlation (in log space) is $r=0.3$, so moderately strong.

\subsection{Linear or Log?}

A first approach to normalizing the number of methods controlling for
size of the project is to make a simple average $WMC = Methods /
Classes$. This is, in fact, how this metric is defined in the
literature~\cite{Chidamber:1991}, assuming uniform complexity of 1 (an
assumption made in several prior studies). This gives us a number
that, in principle, can be used to compare projects independent of
their size. If we have two projects, one with $WMC=3$ and the other
with $WMC=8$, that tells us that these two projects are considerably
different without needing to know any size metric. 

In software ecosystems, a mean of $WMC$ can be calculated for entire
collections of projects by computing the $WMC$ of all the projects in
collection, and then computing the mean of those values. In our
dataset $mean(WMC)=5.15$, which might lead us to conclude that in this
very large Java ecosystem, the average WMC is 5.15.

This value, however, is very misleading, because the distribution of
$WMC$ in the dataset is not normal, but
log-normal. Figure~\ref{fig:wmc} in Appendix shows the $WMC$ distribution in
linear and log scales.

Given this knowledge, a second approach to normalizing for size is to
find the mean and SD of $WMC$ in log scale. In our dataset that is
$mean(WCM)_{log}=1.455$ and $SD(WMC)_{log}=0.63$. This translates to
linear space as $4.28$, with 68\% of values falling within the
interval $[2.28 - 8.00]$, skewed towards the lower end of the
interval.

The first thing to notice is that these two numbers, $mean(WMC)$ and
$mean(WMC)_{log}$ are different, the former being larger than the
latter. That happens because the data is highly right-skewed,
i.e. there are many more smaller values than larger ones. Therefore
the simple mean in linear scale does not capture an important aspect
of the data, its skewness; the mean and SD in log scale do. Another way of
looking at this is that when drawing a data point randomly out of this
dataset, the odds are higher around $4.28$ than around $5.15$.

Even though this is basic statistics, many papers continue to report
summary statistics in linear scale when the data is not normally
distributed in that scale. In general, we must inspect what kind of
distribution our data has and report summary statistics accordingly,
or the reports will be misleading.

\subsection{Non-Linearity}

The above analysis is still missing something important about the
data, namely the findings unveiled by this paper that the number of
methods in a project grows {\em disproportionately} faster with the
number of classes. Therefore {\em normality} takes a different value
depending on the scale of the project. We might conclude that a
project with $WMC=7.9$ ($WMC_{log} = 2.067$), which is on the edge of
the SD interval, might need special attention, and that a project with
$WMC=4.3$ would be perfectly ``normal.'' That may or may not be the
case, depending on the size of that project. In
Section~\ref{sec:modulesize} we found a strong non-linear model given
by:

$Methods = e^{\alpha} Classes^{\beta}$

This equation gives us the {\em norm} of what to expect of $WMC$ in
projects of varying sizes in this dataset. For a project with 10,000
classes, the expectation of the model is that it will have 70,000+
methods, not 42,800 as a simple linear model would predict (i.e. 4.28
* 10,000); so $WMC=7.9$ is what we would expect for a project of this
size. A project with 10,000 classes that shows $WMC=4.3$ would be an
oddity in this ecosystem. If, however, the project has only 25
classes, then $WMC=4.3$ would be expected, but $WMC=7.9$ would be
surprising, in the sense that it is a large deviation from what is
expected of projects of that size. 

Therefore, the proper normalization for size must take this non-linear
relation into account, producing an adjusted ratio that is {\em
  truly} independent of the number of classes:

\begin{equation}
\label{eq4}
%\overset{\text{def}}{=}
WMC_{\beta} = \frac{Methods}{ Classes^{\beta}}
\end{equation}

\begin{figure}
\centering
\includegraphics[width=3in]{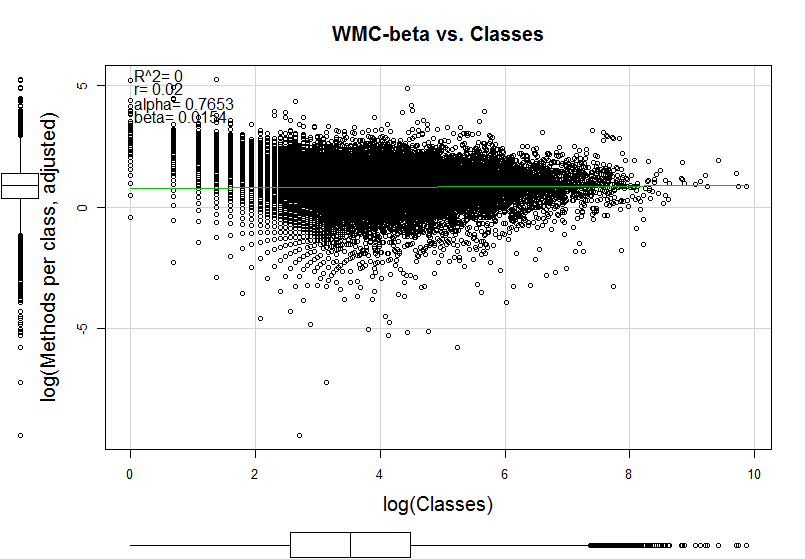}
\caption{Normalization: how $WMC_{\beta}$ grows with the number of classes.}
\label{fig:wmcbeta_classes}
\end{figure}

Figure~\ref{fig:wmcbeta_classes} shows how $WMC_{\beta}$ and size are
{\bf not correlated}, using $\beta=1.1055$, and the parameters from model
1 in the previous section. Pearson correlation between the two
variables is $r << 0.001$, and Spearman correlation is $R=-0.04$.

The parameter $\beta$ has elluded measurement, because it can only be
observed on sufficiently large collections of programs written in the
same language and that, collectively and empirically, define what is
to be expected of programs written in that language. We now have the
means to measure it, as shown in this paper. Therefore, we now have
the knowledge to create updated versions of well-known software
metrics that are truly independent of size and that may (or may not)
carry additional important information about the code that is not
already captured by size metrics. If, for example, high coupling
really is ``bad'', we now have the mathematical knowledge to measure
the size-independent essence of coupling. We plan to investigate the
statistical power of this seamingly small, but critical, adjustment in
future work.

\section{Conclusion}

We have described a quantitative study designed to answer the
question: does the scale of a software system affect the internal
structure of its modules? We have made an important step
into answering this question by performing a statistical analysis of a
very large and varied collection of Java projects. The statistical
significant results in this dataset are strong: there are,
indeed, superlinear effects on some aspects of the modules' internal
structure and composition with other modules. This reinforces the
widely accepted idea that programming-in-the-large carries with it
different concerns that aren't as strongly present for
programming-in-the-small. More importantly, it has tremendous
consequences for software metrics in general. Many of the metrics
proposed in the literature, and that are used widely in IDEs, have
suffered from poor information content for prediction models because
they correlate with the much simpler size metrics. Our paper shows how
this can be corrected.

%% Uncomment later
\acks

This work was supported by National Science Foundation
grants nos. CCF-0725370 and CCF-1018374, and by the DARPA MUSE
program. We would like to thank Pedro Martins for his assistance in
the production of the artifact, and the anonymous reviewers, who made
this a better paper.

% We recommend abbrvnat bibliography style.

\bibliographystyle{abbrvnat}

\bibliography{java-scale}

% The bibliography should be embedded for final submission.

%% \begin{thebibliography}{}
%% \softraggedright

%% \bibitem[Smith et~al.(2009)Smith, Jones]{smith02}
%% P. Q. Smith, and X. Y. Jones. ...reference text...

%% \end{thebibliography}

%\newpage

\appendix

\section{Additional Plots}

\begin{figure}[b]
\centering
\includegraphics[width=3.5in]{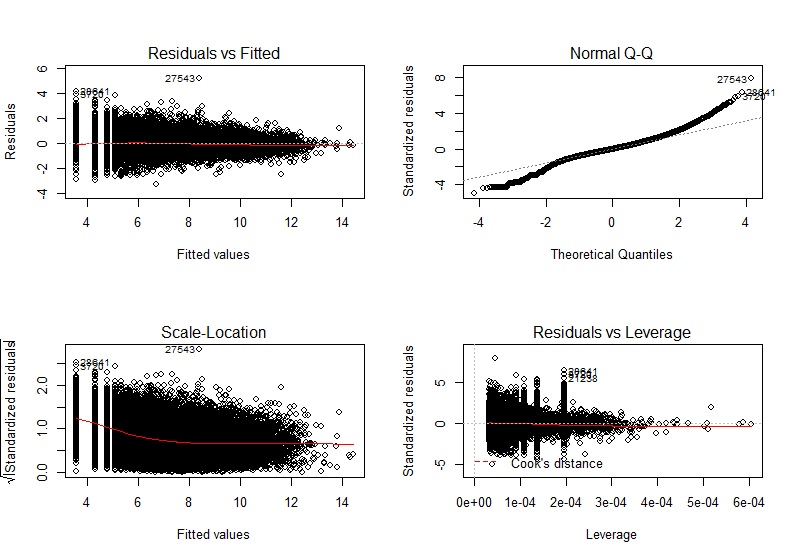}
\caption{Residuals of SLOC $\sim$ Modules.}
\label{fig:res_SLOC_Modules}
\end{figure}

\begin{figure}
\centering
\includegraphics[width=3.5in]{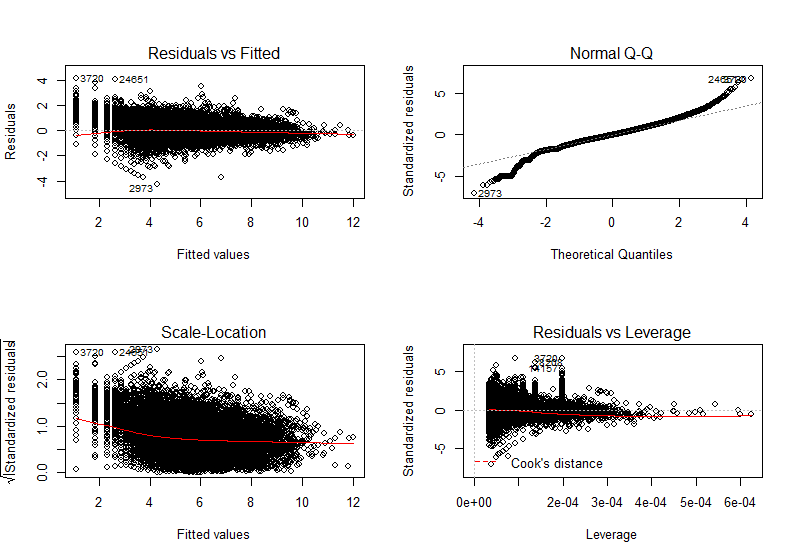}
\caption{Residuals of Methods (in classes) $\sim$ Classes.}
\label{fig:res_Methods_Classes}
\end{figure}

\begin{figure}
\centering
\includegraphics[width=3.5in]{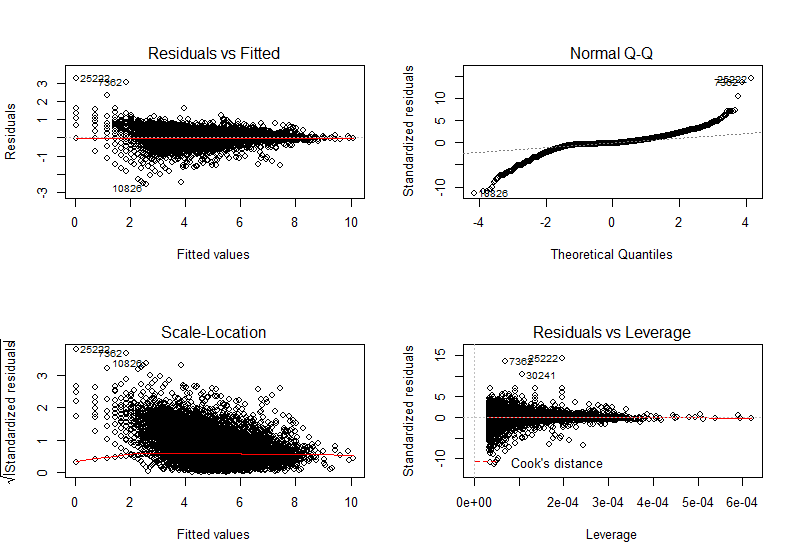}
\caption{Residuals of Constructors $\sim$ Classes. }
\label{fig:res_Constructors_Classes}
\end{figure}

\begin{figure}
\centering
\includegraphics[width=3.5in]{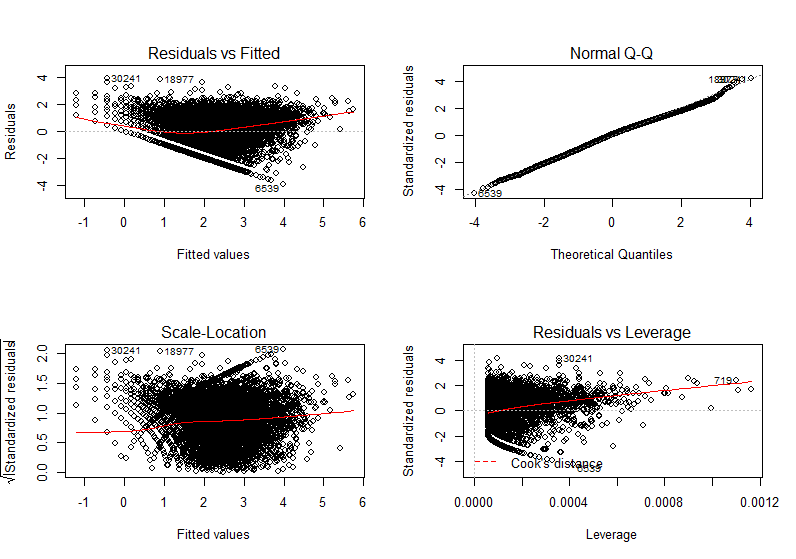}
\caption{Residuals of Interfaces $\sim$ Classes. }
\label{fig:res_Interfaces_Classes}
\end{figure}

\begin{figure}
\centering
\includegraphics[width=3.5in]{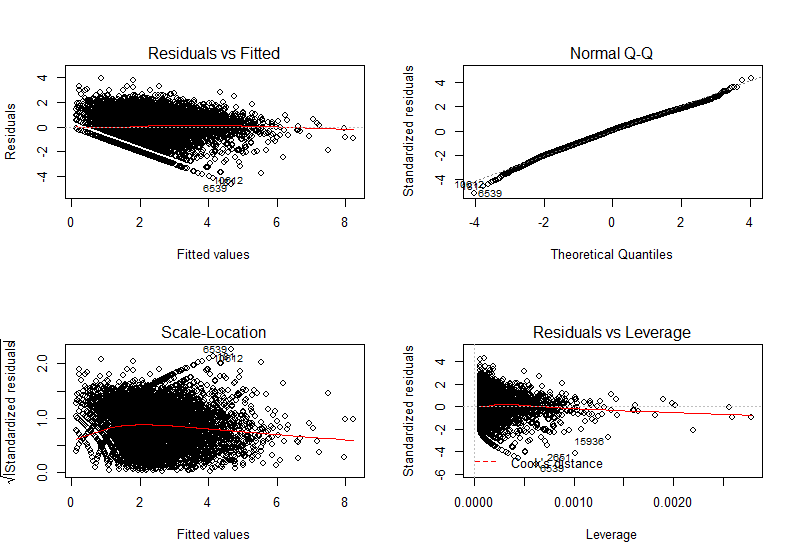}
\caption{Residuals of Interfaces $\sim$ Classes. Classes in $log^{2}$ scale.}
\label{fig:res_Interfaces_Classes2}
\end{figure}

\begin{figure}
\centering
\includegraphics[width=3.5in]{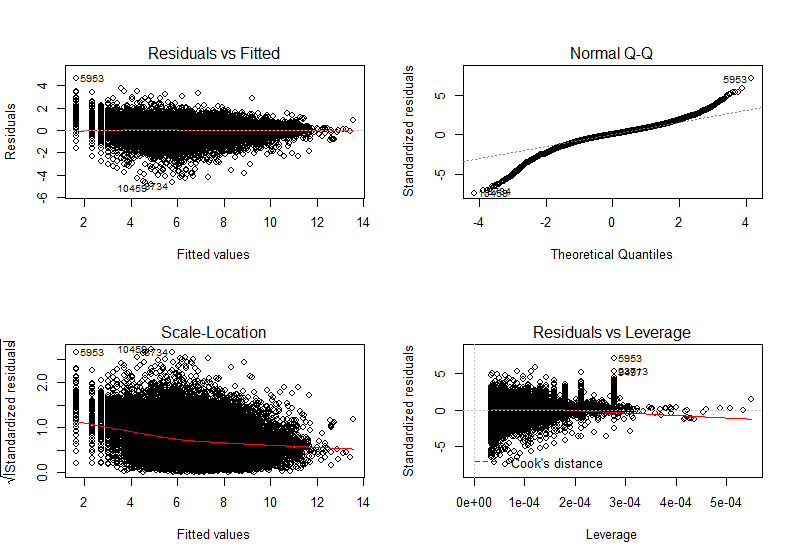}
\caption{Residuals of Calls $\sim$ Methods. }
\label{fig:res_Calls_Methods}
\end{figure}

\begin{figure}
\centering
\includegraphics[width=3.5in]{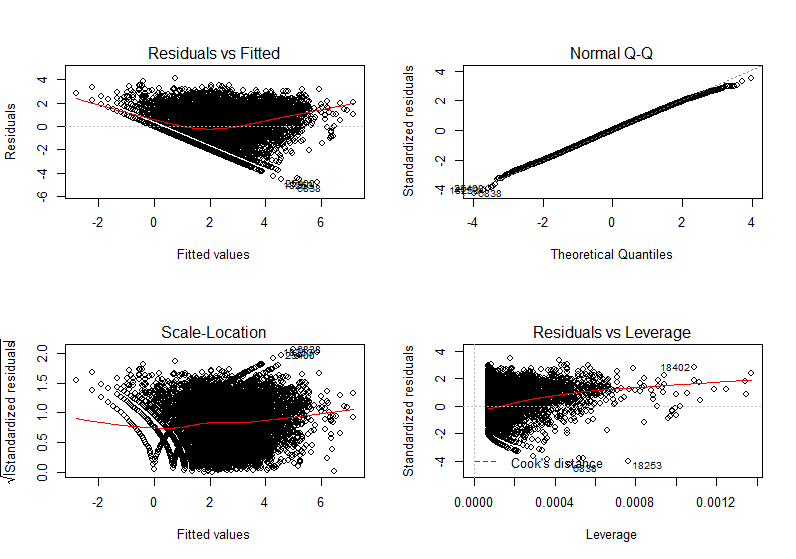}
\caption{Residuals of Instanceof statements $\sim$ Methods.}
\label{fig:res_Instanceof_Methods}
\end{figure}

\begin{figure}
\centering
\includegraphics[width=3.5in]{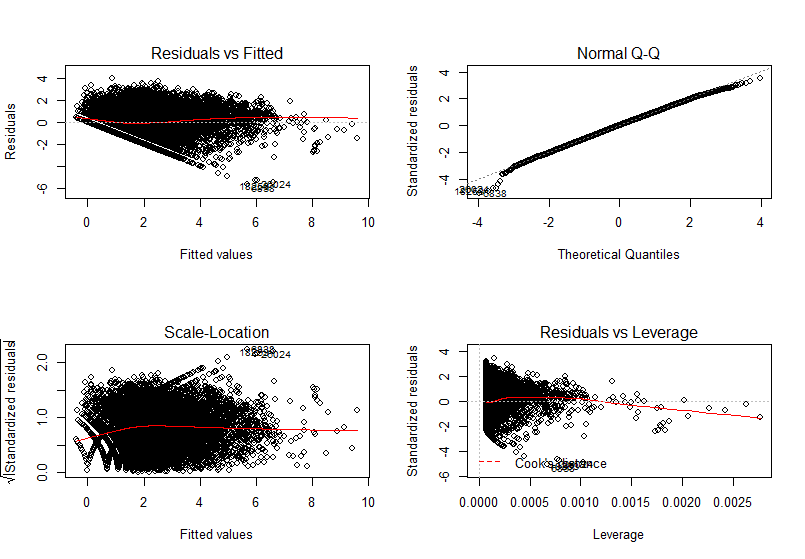}
\caption{Residuals of Instanceof statements $\sim$ Methods. Methods in $log^{2}$ scale.}
\label{fig:res_Instanceof_Methods2}
\end{figure}

\begin{figure}
\centering
\includegraphics[width=3.5in]{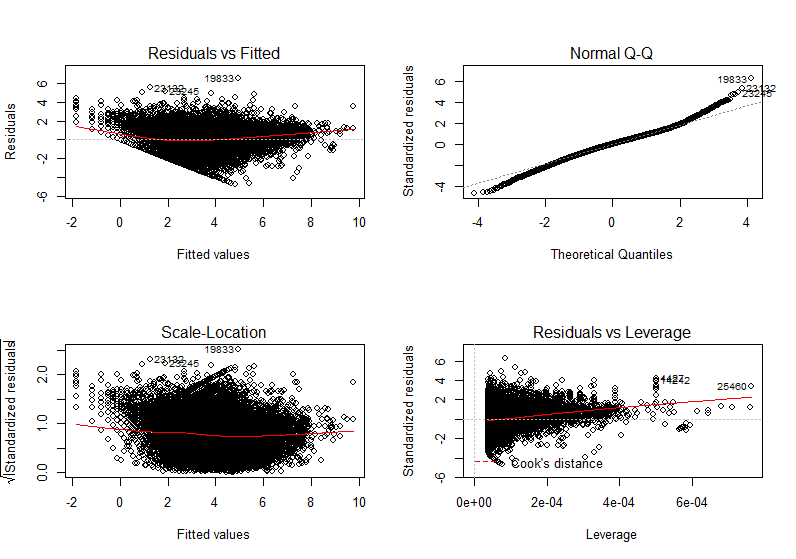}
\caption{Residuals of Casts $\sim$ Methods.}
\label{fig:res_Casts_Methods}
\end{figure}

\begin{figure}
\centering
\includegraphics[width=3.5in]{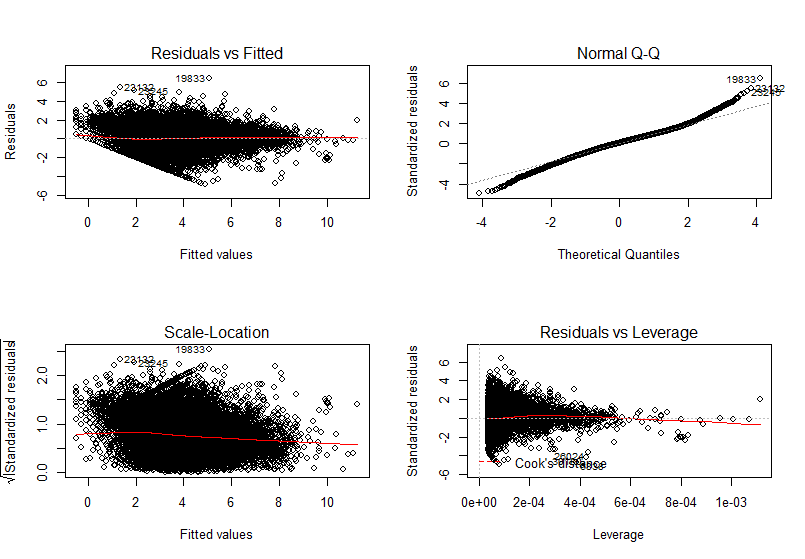}
\caption{Residuals of Casts $\sim$ Methods. Methods in $log^{1.4}$ scale.}
\label{fig:res_Casts_Methods2}
\end{figure}

\begin{figure}
\centering
\includegraphics[width=3.5in]{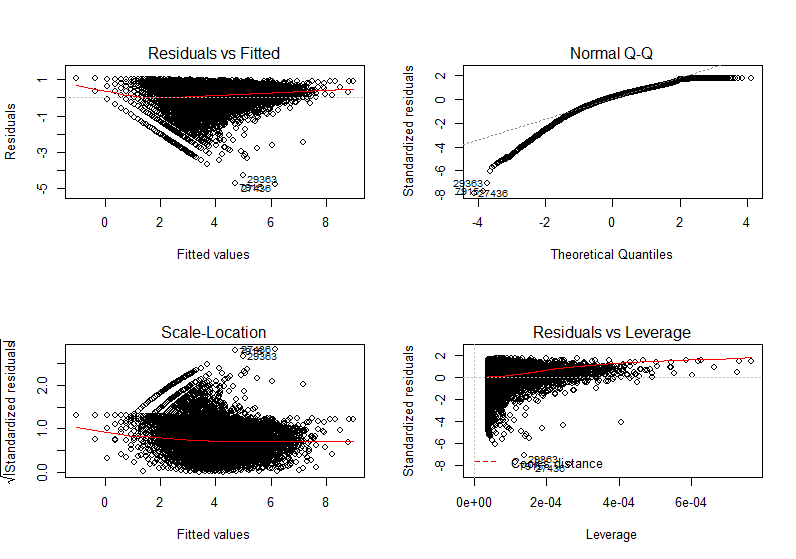}
\caption{Residuals of classes defined using inheritance (DUI) $\sim$
  Classes. }
\label{fig:res_DUI_Classes}
\end{figure}

\begin{figure}
\centering
\includegraphics[width=3.5in]{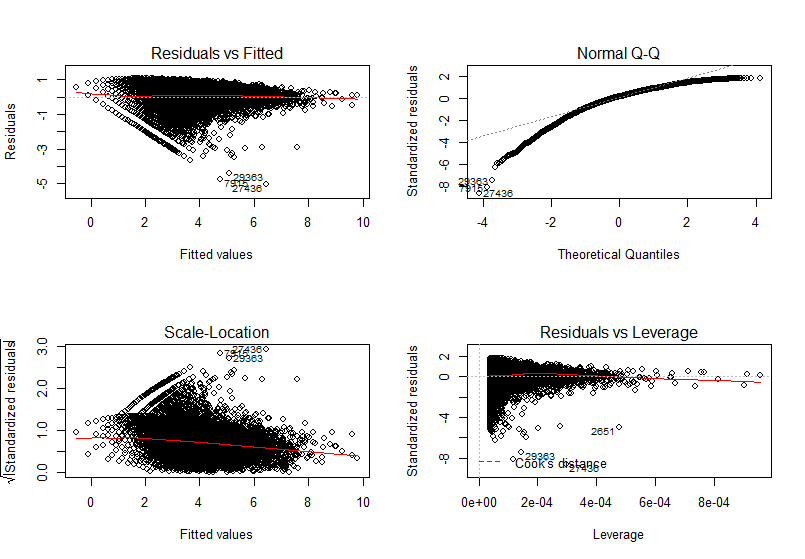}
\caption{Residuals of classes defined using inheritance (DUI) $\sim$
  Classes. Classes in $log^{1.2}$ scale.}
\label{fig:res_DUI_Classes2}
\end{figure}

\begin{figure}
\centering
\includegraphics[width=3.5in]{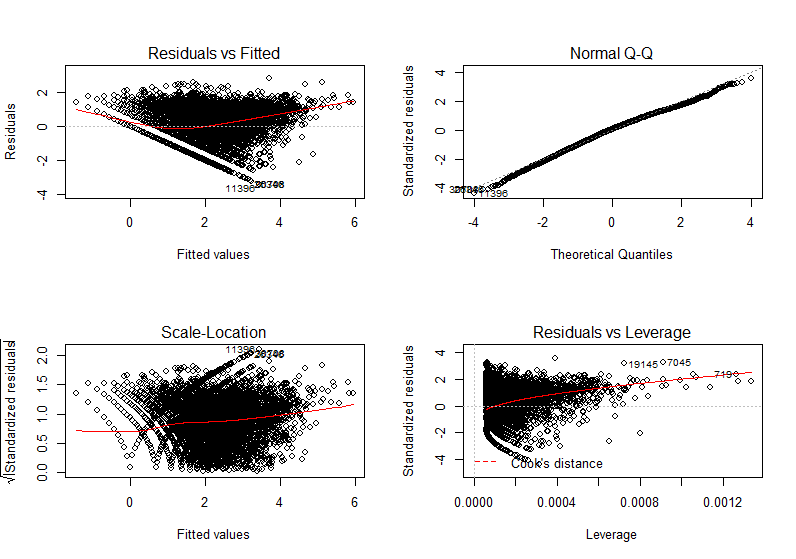}
\caption{Residuals of classes inherited from (IF) $\sim$ Classes.}
\label{fig:res_IF_Classes}
\end{figure}

\begin{figure}
\centering
\includegraphics[width=3.5in]{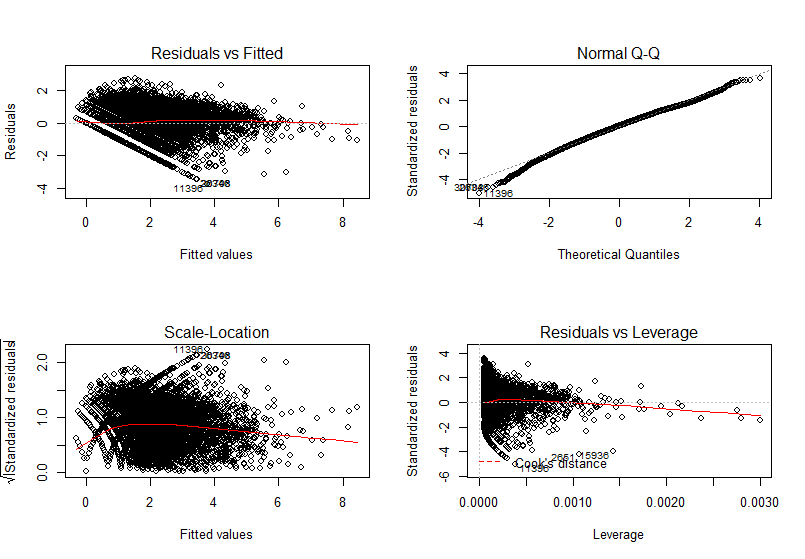}
\caption{Residuals of classes inherited from (IF) $\sim$ Classes. Classes in $log^{2}$ scale.}
\label{fig:res_IF_Classes2}
\end{figure}

%\clearpage
\begin{figure}
\centering
\includegraphics[width=3.5in]{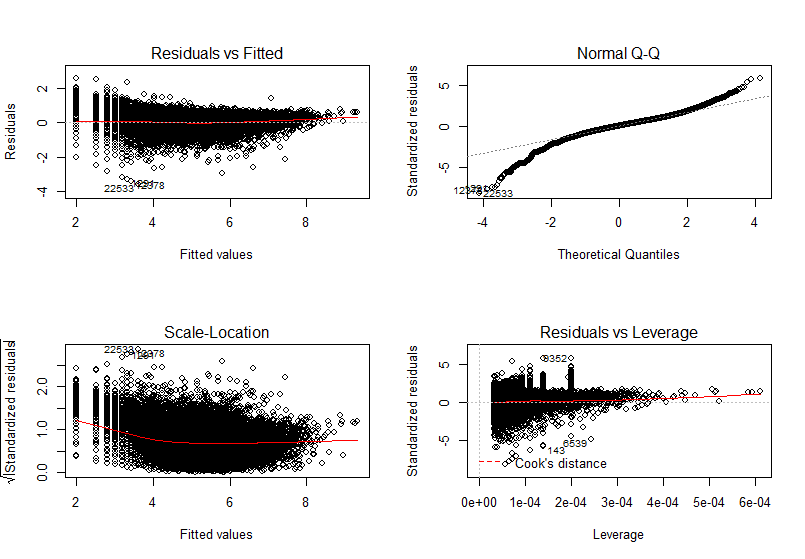}
\caption{Residuals of used modules $\sim$ Declared modules. Classes in $log^{2}$ scale.}
\label{fig:res_Used_Modules}
\end{figure}

\begin{figure}
\centering
\includegraphics[width=3.5in]{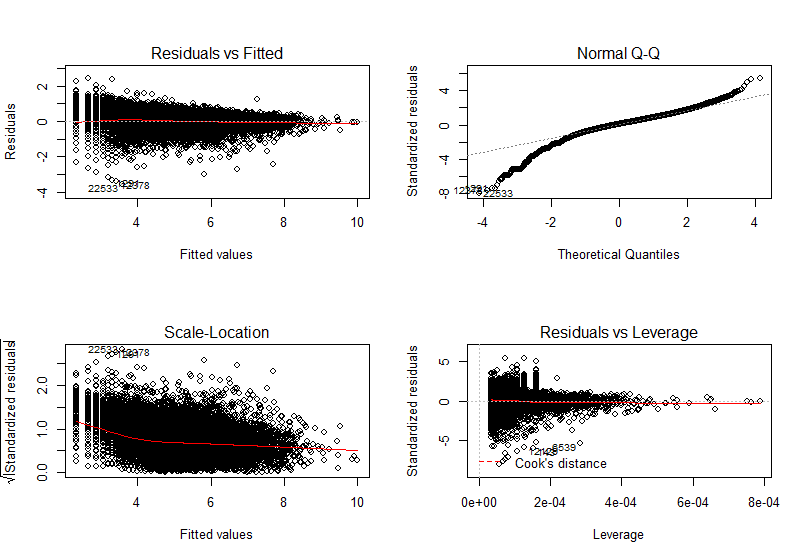}
\caption{Residuals of used modules $\sim$ Declared modules. Classes in $log^{2}$ scale.}
\label{fig:res_Used_Modules2}
\end{figure}
\clearpage

\begin{figure}
\centering
\includegraphics[width=3.5in]{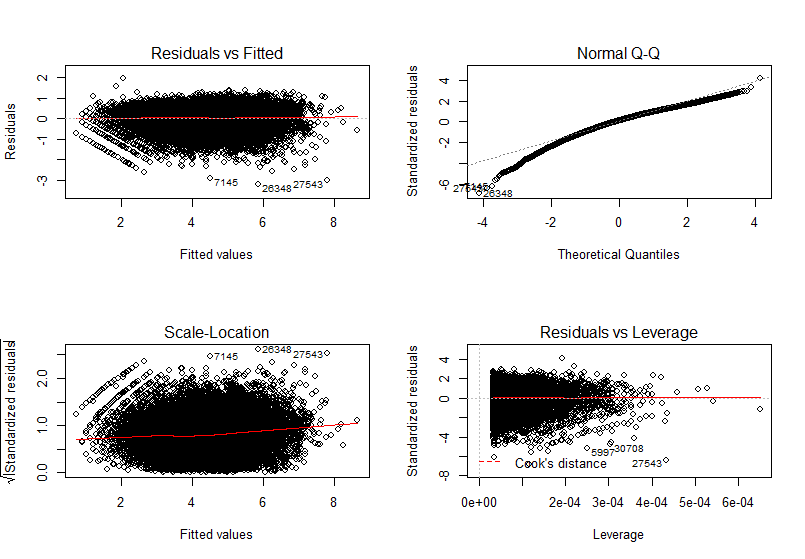}
\caption{Residuals of efferent coupling $\sim$ SLOC. }
\label{fig:res_Coupling_SLOC}
\end{figure}

\begin{figure}
\centering
\includegraphics[width=3.5in]{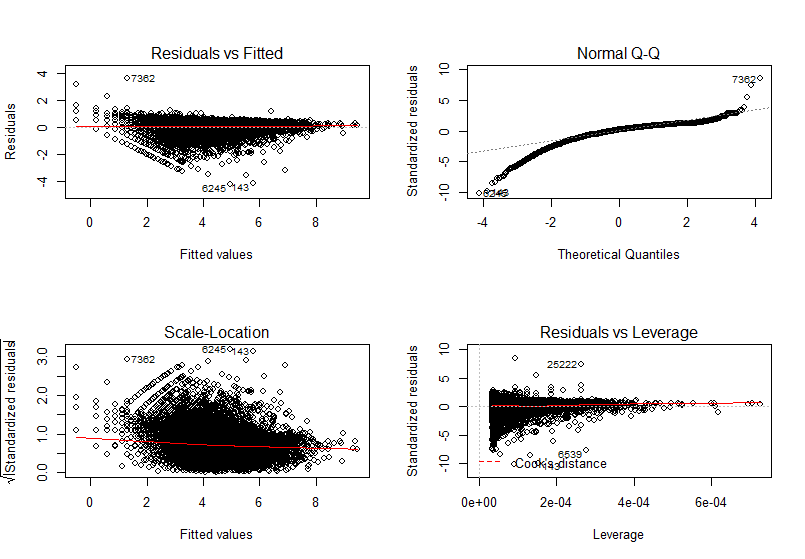}
\caption{Residuals of used internal modules $\sim$ Declared modules. }
\label{fig:res_Internal_Modules}
\end{figure}

\begin{figure}
\centering
\includegraphics[width=3.5in]{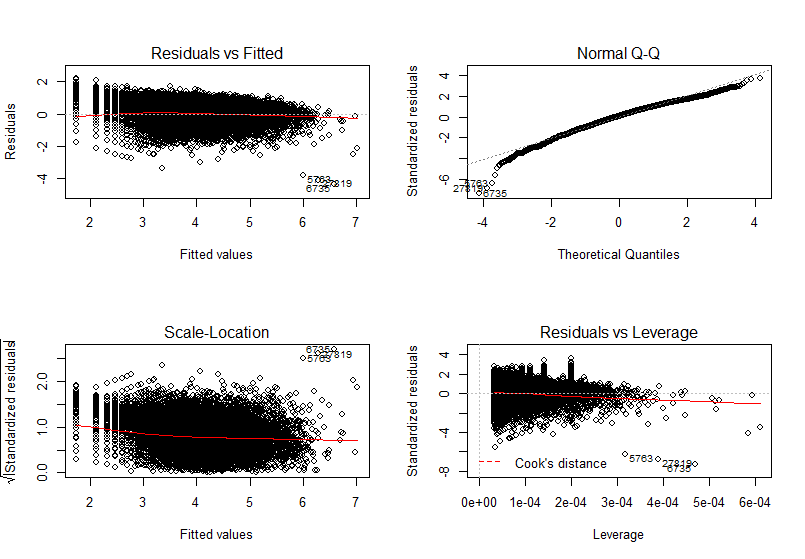}
\caption{Residuals of used JDK modules $\sim$ Declared modules. }
\label{fig:res_JDK_Modules}
\end{figure}

\begin{figure}
\centering
\includegraphics[width=3.5in]{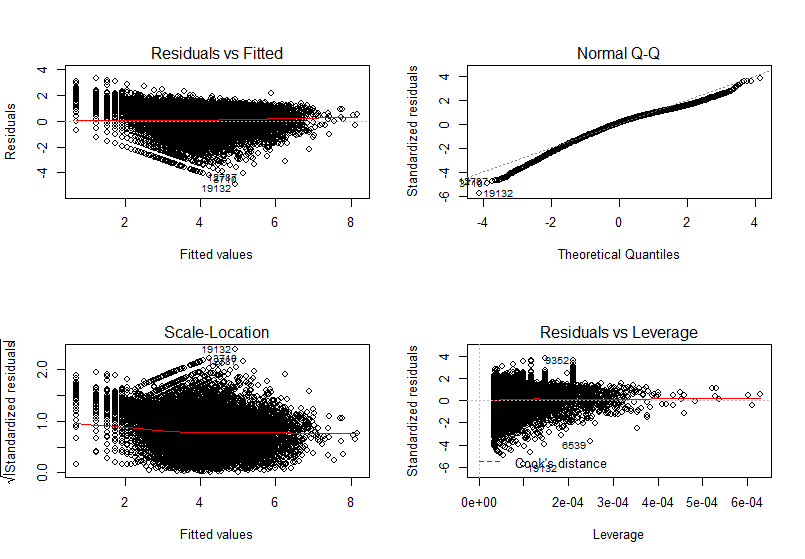}
\caption{Residuals of used external modules $\sim$ Declared modules. }
\label{fig:res_External_Modules}
\end{figure}

\begin{figure}
\centering
\includegraphics[width=3.5in]{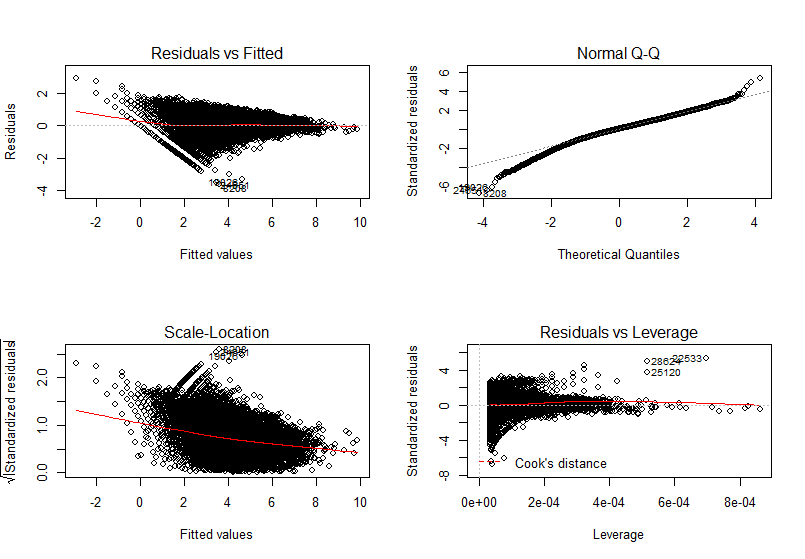}
\caption{Residuals of used internal modules $\sim$ Total used modules. }
\label{fig:res_Internal_Total}
\end{figure}

\begin{figure}
\centering
\includegraphics[width=3.5in]{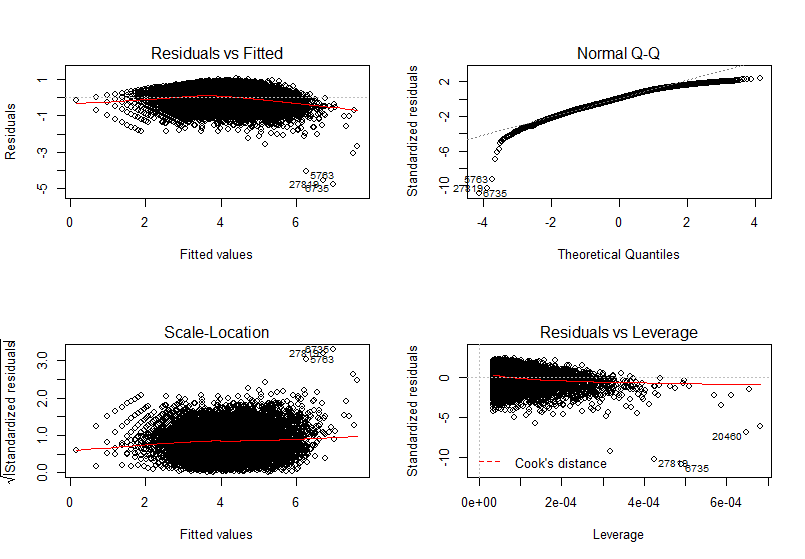}
\caption{Residuals of used JDK modules $\sim$ Total used modules. }
\label{fig:res_JDK_Total}
\end{figure}

\begin{figure}
\centering
\includegraphics[width=3.5in]{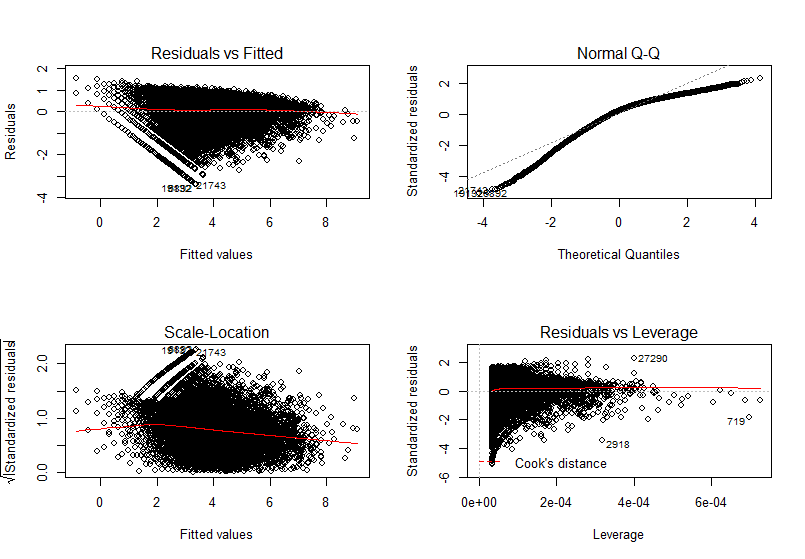}
\caption{Residuals of used external modules $\sim$ Total used modules. }
\label{fig:res_External_Total}
\end{figure}

\begin{figure}
\centering
\includegraphics[width=3.5in]{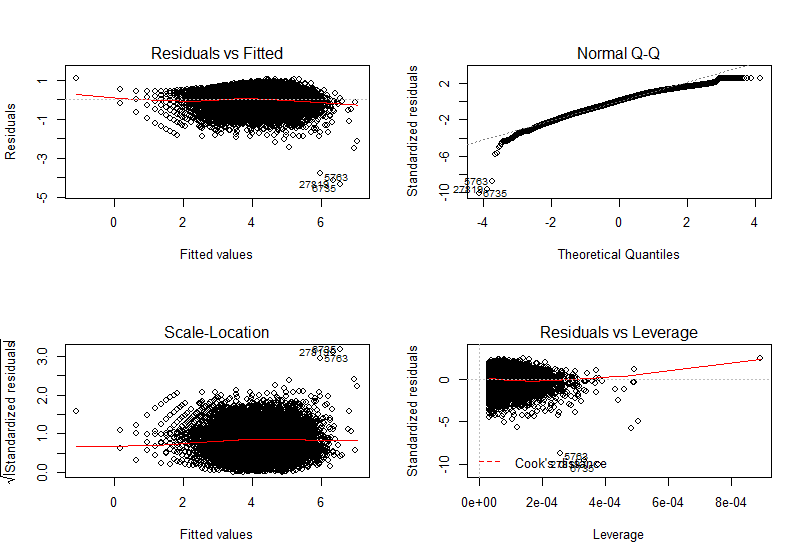}
\caption{Residuals of used internal modules $\sim$ Total used modules. RLM.}
\label{fig:res_Internal_Total2}
\end{figure}

\begin{figure}
\centering
\includegraphics[width=3.5in]{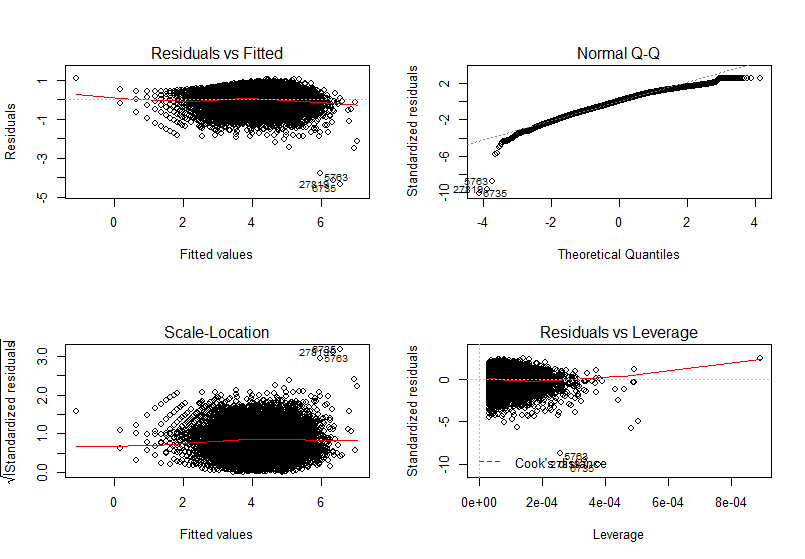}
\caption{Residuals of used JDK modules $\sim$ Total used modules. RLM.}
\label{fig:res_JDK_Total2}
\end{figure}

\begin{figure}
\centering
\includegraphics[width=3.5in]{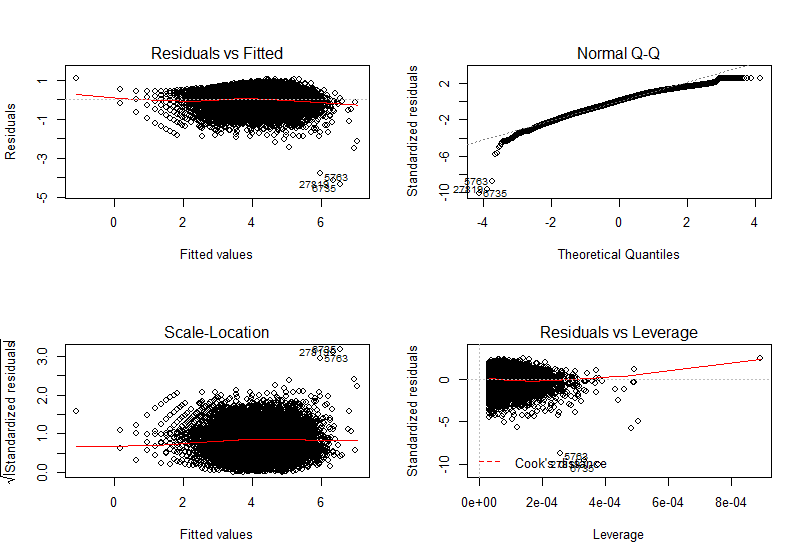}
\caption{Residuals of used external modules $\sim$ Total used modules. RLM.}
\label{fig:res_External_Total2}
\end{figure}

\begin{figure}
\centering
\includegraphics[width=3in]{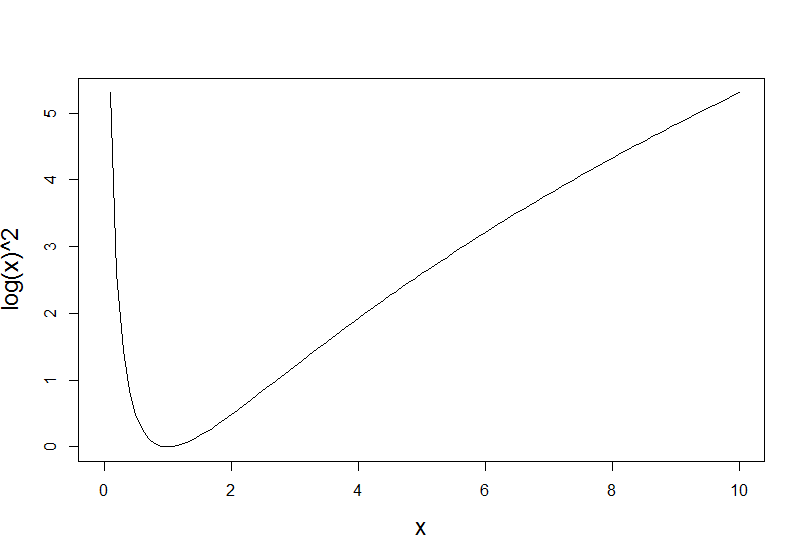}
\caption{Generic $log(x)^{2}$ function.}
\label{fig:log_squared}
\end{figure}

\begin{figure}
\centering
\includegraphics[width=1.5in]{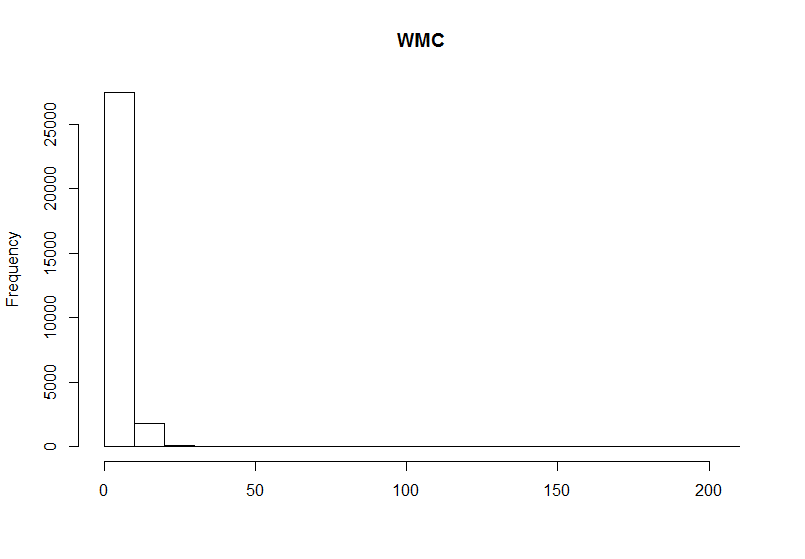}
\includegraphics[width=1.5in]{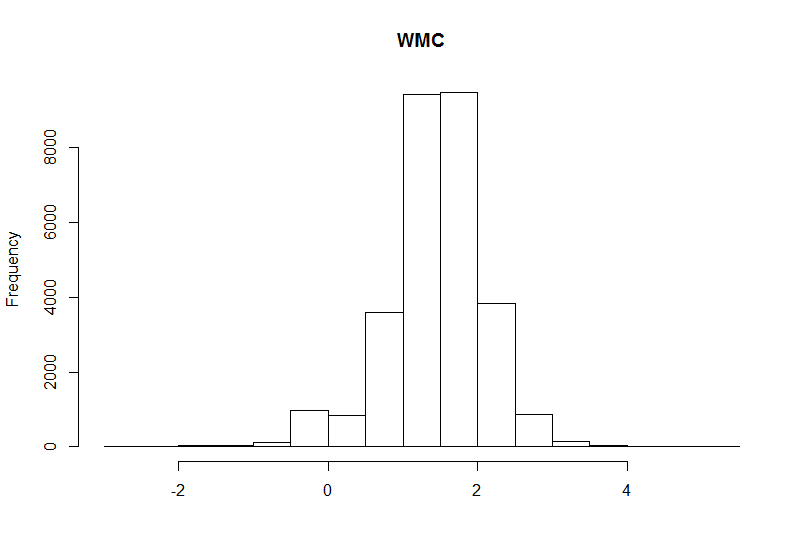}
\caption{Histogram of WMC. Left: Linear scale. Right: Log scale.}
\label{fig:wmc}
\end{figure}

\end{document}